\documentclass[12pt,english]{oxarticle}

\usepackage{graphicx, float, array, xspace, amscd, amsmath, amsthm, amssymb, latexsym, bbm}
\usepackage[sort&compress, comma, square, numbers]{natbib}

\font\eightrm=cmr8 at 8pt

\newcommand{\be}{\begin{equation}}
\newcommand{\ee}{\end{equation}}
\newcommand{\mbb}{\mathbb}
\newcommand{\mc}{\mathcal}

\numberwithin{equation}{section}
\proofmodefalse
\begin{document}
\title{Axion decay constants away from the lamppost}
\author{Joseph P. Conlon, Sven Krippendorf\\
{\it \small Rudolf Peierls Centre for Theoretical Physics, University of Oxford,}\\[-0.1cm] {\it \small 1 Keble Road, Oxford, OX1 3NP, England}}
\maketitle
\begin{abstract}
It is unknown whether a bound on axion field ranges exists within quantum gravity. We study axion field ranges using extended supersymmetry, in particular allowing an analysis within strongly coupled regions of moduli space.
 We apply this strategy to Calabi-Yau compactifications with one and two K\"ahler moduli.
We relate the maximally allowable decay constant to geometric properties of the underlying Calabi-Yau geometry.
In all examples we find a maximal field range close to the reduced Planck mass (with the largest field range being 3.25 $M_P$). On this perspective,
field ranges relate to the intersection and instanton numbers of the underlying Calabi-Yau geometry.
\end{abstract}
\newpage
\tableofcontents
\section{Introduction}

In field theory axions provide excellent candidates for models of large field inflation. The simplest example is natural inflation~\cite{Freese:1990rb}, where
to provide sizeable tensor modes the associated axion decay constant is required to be trans-Planckian.
In the case of two (or more) axions, the UV theory does not have to include trans-Planckian decay constants, but
an effective trans-Planckian decay constant can emerge through either an
appropriate alignment of decay constants~\cite{Kim:2004rp} or the collective behaviour of many axions (N-flation)~\cite{Dimopoulos:2005ac}.

It is a non-trivial question whether the embedding of such effective field theory models can be achieved within a theory of quantum gravity.
This is also not just a theoretical question; trans-Planckian field excursions during inflation can generate
levels of tensor modes that would be observable by current and upcoming experiments.
It therefore represents an ideal topic for string phenomenology, as it both has
observational content and requires a UV-complete theory for a sensible analysis.

Over the last few years, there has been much work targeting this question.
One direction has been the attempt to construct direct models of large field inflation~\cite{Kim:2004rp,Dimopoulos:2005ac,Silverstein:2008sg,McAllister:2008hb,Cicoli:2008gp,Kaloper:2008fb,Berg:2009tg,Palti:2014kza,Marchesano:2014mla,Blumenhagen:2014gta,Hebecker:2014eua,Ibanez:2014kia,Ben-Dayan:2014lca,Baumann:2014nda,Westphal:2015eva} and, by showing that
these are under control, aiming to produce explicit examples of trans-Planckian field excursions in string theory.
On the other hand, many arguments have been put forward, both in the context of critical analysis of explicit models and also through more general grounds,
that suggest problems arise with either consistency or backreaction
once field ranges become trans-Planckian. These arguments include field ranges~\cite{Banks:2003sx,Baumann:2006cd}, entropy bounds \cite{Conlon:2012tz} (for a contrary view see \cite{Kaloper:2015jcz}), backreaction~\cite{Conlon:2011qp,Palti:2015xra} and the use of the
weak gravity conjecture \cite{ArkaniHamed:2006dz, delaFuente:2014aca,Brown:2015iha,Montero:2015ofa,Bachlechner:2015qja,Hebecker:2015rya,Brown:2015lia,Rudelius:2015xta,Junghans:2015hba,Heidenreich:2015wga, Heidenreich:2015nta,Kooner:2015rza,Hebecker:2015zss}.

While these problems take different forms in different corners of string theory, their repeated occurrence
has led many to contemplate the possibility that deep reasons of quantum gravity may
restrict the axionic field range to be sub-Planckian
(although it is left open whether this would mean strictly below the Planck mass or some multiple of the Planck mass $f_a< (2\pi)^n M_{P}$).

How best to analyse this further?
One clear difficulty in studying field ranges in the context of inflationary models in string theory is that these necessarily require
compactifications with supersymmetry broken at a high scale. There are typically also many (hundreds) of moduli present (see for instance string theoretic attempts to realise N-flation~\cite{1401.2579}), and it is essential that none of these
have any unstable runaway directions during inflation. Given that models are often complex and with many moving parts,
it is not easy to guarantee control of the computations, and to ensure that there really are no instabilities
present.\footnote{As a minimal point, an absence of instabilities in four-dimensional effective field theory is no guarantee of an absence of instabilities
in the higher-dimensional theory~\cite{Witten:1981gj}.}

In this paper, we take a different approach. If deep considerations of quantum gravity ensure a bound on axion field ranges in string theory, this bound should also
apply for models with unbroken and extended supersymmetry. Extended supersymmetry also allows for exact results, which can be analysed even in
normally inaccessible strong coupling regions dominated by non-perturbative physics.
It also allows for precise study of the physics that truncates the axion decay constant, and for a determination of the maximum axion decay constant, including factors of $\pi$,
arising in the various examples studied.

There are certainly disadvantages of this approach. The study of a number of distinct examples does not imply any general proof. The study of decay constants in
models with unbroken supersymmetry and vanishing potential also does not say anything about the feasibility of developing an effective trans-Planckian decay constant using (for example) alignment of the potential for
many axions with individually sub-Planckian decay constants.

Nonetheless, we still think it worthwhile to study the precise decay constants that arise in situations where there is full calculational control. While this has been considered before in \cite{Banks:2003sx}, that paper was concerned only
with approximate relations and indeed absolved itself from factors of $16 \pi^2$. Given that such factors would have a crucial impact of the observability of tensor modes, we
think a quantitative study is important.

To this end, we shall study axionic field ranges for Calabi-Yau compactifications of type IIA string theory,
using the exact results available for the K\"ahler moduli space that come from mirror symmetry applied to the
complex structure moduli space of type IIB string theory compactified on the mirror manifold.

The paper is organised as follows.
In Section~\ref{sec:dilatonaxionsystem} we discuss as a warm-up example the familiar dilaton-axion system,
 explaining how the axion field range is bounded by the $SL(2, \mbb{Z})$ duality symmetry.
 We then move in Section~\ref{sec:onemodulus} to the case of Calabi-Yau manifolds with one K\"ahler modulus, analysing a variety of examples
 and determining the maximal axion field range in each one.
We extend these considerations to models with two K\"ahler moduli in Section~\ref{sec:twomoduli}.
We make some comments on the connection to field theory models of large field inflation using axions in Section~\ref{sec:connection},
before concluding in Section~\ref{sec:conclusion}.

\section{The Dilaton-Axion System}
\label{sec:dilatonaxionsystem}

We start with a familiar and baby example illustrating the physics that will recur throughout this paper.
 By an `axion', we denote in this paper a real scalar field $a$ with a periodicity
$$
a \equiv a + 2 \pi f_a~,
$$
such that the field configurations $\langle a \rangle = \theta$ and $\langle a \rangle = \theta + 2 \pi f_a$ represent identical states in the Hilbert space,
and such that all intermediary values of $\langle a \rangle$ are inequivalent. We are interested in the maximal value that $f_a$ can attain in string theory.

Axions can arise in many different contexts within string theory. In this paper we study the values of $f_a$
attained in regimes that are out of perturbative control, and to this end we will consider
systems with exact supersymmetry. As is well known, extended supersymmetry highly constrains the effective actions and in some cases allows for exact
solutions within the strongly coupled parts of moduli space.

One of the most universal axions occurring in string theory is that of the dilaton-axion system. We
work in the context of type IIB string theory compactified on a Calabi-Yau manifold, preserving $\mc{N}=2$ supersymmetry.
The classical K\"ahler potential for the dilaton-axion system is
\be
\label{Kdilatonaxion}
K = - M_P^2 \ln \left( S + \bar{S} \right),
\ee
which descends from the ten-dimensional IIB supergravity action, with $S = c_0 + \frac{i}{g_s} \equiv s_R + I s_I$ a combination of the RR 0-form axion and the string coupling.
The axionic identification is $S \equiv S + 1$.
 This K\"ahler potential gives rise to the Lagrangian kinetic terms
\be
\frac{M_P^2}{4 s_I^2} \partial_{\mu} s_I \partial_{\mu} s_I + \frac{M_P^2}{4 s_I^2} \partial_{\mu} s_R \partial_{\mu} s_R~.
\ee
At large values of $s_I$, we are in a weakly coupled regime where the field $s_R$ has the classical axion behaviour.
The physical size of the axion field range is the length of the displacement $s_R \to s_R + 1$, namely
\be
\label{axiondilatonfieldrange}
2 \pi f_a = \frac{g_s M_P}{\sqrt{2}}~.
\ee
In the `classical' regime of large ${\rm{Im}}(S)$ and weak $g_s$, we have the familiar notion of an axion.
In this region the decay constant is also clearly sub-Planckian. As $g_s$ increases,
the length of the displacement $s_R \to s_R + 1$ also increases towards a Planckian field displacement.

What happens as $g_s$ increases towards (and beyond) unity? What does \emph{not} happen is that large $g_s$-corrections, either perturbative
or non-perturbative, destroy the validity of the effective action of equation (\ref{Kdilatonaxion}).
While it is true that the 10-dimensional action receives corrections in both $\alpha'$ and $g_s$,
the $g_s$-corrections -- both perturbative and non-perturbative D-instanton corrections --
are subsumed into $\alpha'$-corrections, and so first arise only at order $\alpha'^{3}$. In particular, there are no $g_s$-corrections that are tree-level in
$\alpha'$.

The relevance of this is that all $g_s$-corrections are suppressed by factors of volume, and so
provided the extra-dimensional volume is also large,
the classical metric of (\ref{Kdilatonaxion}) remains an accurate
description of the dilaton-axion moduli space even at large values of the string coupling (for example, see ~\cite{RoblesLlana:2006ez}).

\begin{figure}
\begin{center}
\includegraphics[height=0.38\textheight]{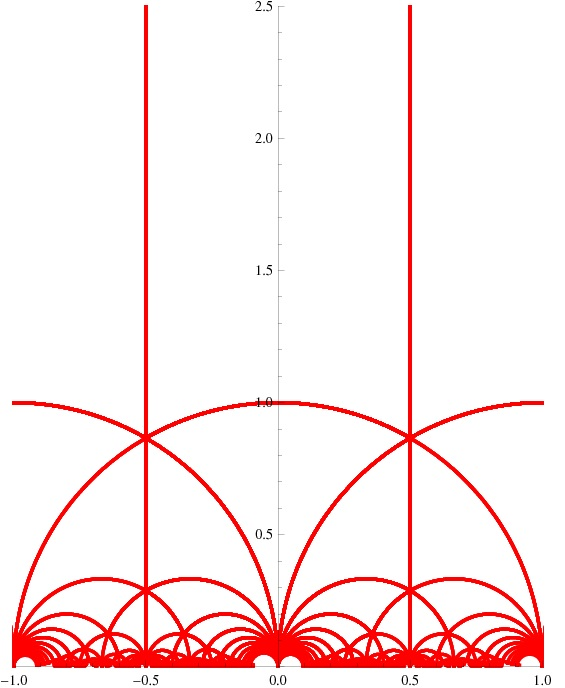}
\end{center}
\caption{Fundamental domain under the discrete $SL(2, \mbb{Z})$ of the dilaton-axion system.}
\label{SL2Zregion}
\end{figure}

Naively, as we enter the region $g_s > 1$ the axion decay constant becomes trans-Planckian.
However, this is not really true, as
the moduli space of the dilaton-axion multiplet $S$ is controlled by the $SL(2, \mbb{Z})$ symmetry,
generated not only by the axionic shift symmetry $S \equiv S + 1$ but also by $S \equiv -1/S$.
The discrete $SL(2, \mbb{Z})$ symmetry restricts the fundamental region of the dilaton-axion system as shown in Figure~\ref{SL2Zregion}.
While without the action of $SL(2, \mbb{Z})$, the axionic decay constant would indeed grow indefinitely,
we see that this conventional interpretation can only be sustained until we reach $\rm{Im}(S) = 1$.
Below this, the axionic `path' can be reinterpreted as simply a different route within the fundamental region.
The truncation to the fundamental region then provides a maximal value for the axion decay constant of $2 \pi f_a = \frac{M_P}{\sqrt{2}}$.

We note that it is the intrinsically stringy physics of the $SL(2, \mbb{Z})$ duality group that is responsible for cutting off the moduli space
here, and thereby capping the axion decay constant.
We also note that this is despite the fact that the effective action remains good down to arbitrarily small values of $\rm{Im}(S)$,
even though formally $g_s \gg 1$ in this region.

This example is hopefully familiar, but its key elements will recur later.
These elements are the presence of an axionic field in the 4-d effective theory, that
in the classical limit of large volume and weak coupling has a highly sub-Planckian decay constant.
As we move into the `stringy' regime of moduli space, the decay constant grows towards a Planckian value, and would classically become
trans-Planckian. However intrinsically stringy physics, and in particular duality groups that restrict the moduli space to a fundamental region,
caps the field range at a finite maximum value close to the Planck scale.

\section{One Parameter models}
\label{sec:onemodulus}

We now move onto examples of Calabi-Yau manifolds with one K\"ahler modulus for which the exact prepotential has been calculated in the literature.

We study here type IIA string theory compactified on a Calabi-Yau 3-fold. In this case, the K\"ahler moduli fall into vector multiplets and the complex structure moduli into
hypermultiplets (in type IIB compactifications,
it is the complex structure moduli that fall into vector multiplets while the K\"ahler moduli belong in hypermultiplets).
For both type IIA and type~IIB cases the dilaton falls in a
hypermultiplet.

As is well known, the product nature of $\mathcal{N}=2$ supersymmetry implies that the hypermultiplet and vector multiplet moduli spaces form a direct product.
In particular, while the type~IIA K\"ahler moduli space receives $\alpha'$-corrections, there are no $g_s$-corrections. Furthermore, the form of these corrections can be determined via
mirror symmetry techniques, allowing the structure of the IIA K\"ahler moduli space to be solved -- even at large $g_s$ -- at strong coupling in the $\alpha'$ expansion.

In classical type IIA compactifications, the K\"ahler moduli correspond
to the volumes of 2-cycles, with their axionic parts corresponding to the reduction of the NS-NS 2-form $B_2$ on the cycle.
At large volumes and weak coupling, these provide good examples of axions. As for the dilaton-axion system, we can determine the classical axion field range,
and then using the mirror symmetry computations we can follow the field range into small volumes.

In this section, we
do this for a number of one-modulus Calabi-Yau manifolds, and thereby study the axionic field range in strong coupling regions where calculational
control is normally inaccessible. We take the relevant data for these from a variety of both classic and more recent papers on mirror symmetry.

\subsection{The Quintic}

The canonical example of mirror symmetry is the quintic.
We compactify type IIA string theory on the
quintic hypersurface of $\mathbb{CP}^4$ with $(h^{1,1}, h^{2,1}) = (1,101)$ that is described by
the following degree five equation
\begin{equation}
p=\sum_i x_i^5-\psi_5( x_1,  x_2,  x_3,  x_4,  x_5) = 0~.
\end{equation}
As is well known, the K\"ahler moduli space on this manifold can be obtained from the classical complex structure moduli space for $\psi$ on the mirror manifold
with $(h^{1,1}, h^{2,1}) = (101,1)$, described by the equation
\be
p = \sum_i x_i^5 - \psi x_1 x_2 x_3 x_4 x_5 = 0~,
\ee
and obtained from the original case by a quotient under the action of a discrete $\mbb{Z}_5^3$ symmetry.

The resulting prepotential has been calculated in~\cite{Candelas:1990rm} and is given by\footnote{A summary of our conventions is given in Appendix~\ref{sec:appconventions}.}
\begin{equation}
F=-\frac{5}{6}T^3-\frac{11}{4}T^2+\frac{25}{12}T+\frac{\chi(M) i\zeta{(3)}}{2 (2 \pi)^3}-\frac{1}{(2\pi i)^3} \sum_{k=1}^\infty n_k~{\rm Li}_3{(e^{2\pi i k T})}~,
\end{equation}
where $T = a + it$, the Euler number is $\chi(M) = -200$ and the leading $n_k$ coefficients are given by:
\begin{equation}
\begin{array}{c|| l | l | l | l | l | l}
k & 1 & 2 & 3 & 4 & 5 & 6 \\ \hline
n_k & 2875 & 609250 & 317206375 & 242467530000 & 229305888887625 & 248249742118022000
\end{array}
\end{equation}
The classical (tree-level in $\alpha'$) K\"ahler potential is
\be
K = - 3 \ln \left( -i (T - \bar{T}) \right) + {\rm constant}.
\ee
The axion periodicity $a \equiv a + 2 \pi f_a$ is set by
$$
2 \pi f_a = \sqrt{\frac{3}{2}} \frac{M_P}{t},
$$
where $t$ is the volume of the 2-cycle measured in units of $l_s = 2 \pi \sqrt{\alpha'}$, with the classical Calabi-Yau volume (in the same units) given by $\mc{V} = (5/6)t^3$.
As for the dilaton-axion system, in the limit of large $t$ the decay constant is clearly sub-Planckian, approaching Planckian values as $t$ tends towards 1.
Our interest is in a controlled analysis of the decay constant as $t$ approaches 1.

\begin{figure}
\begin{center}
\includegraphics[width=0.5\textwidth]{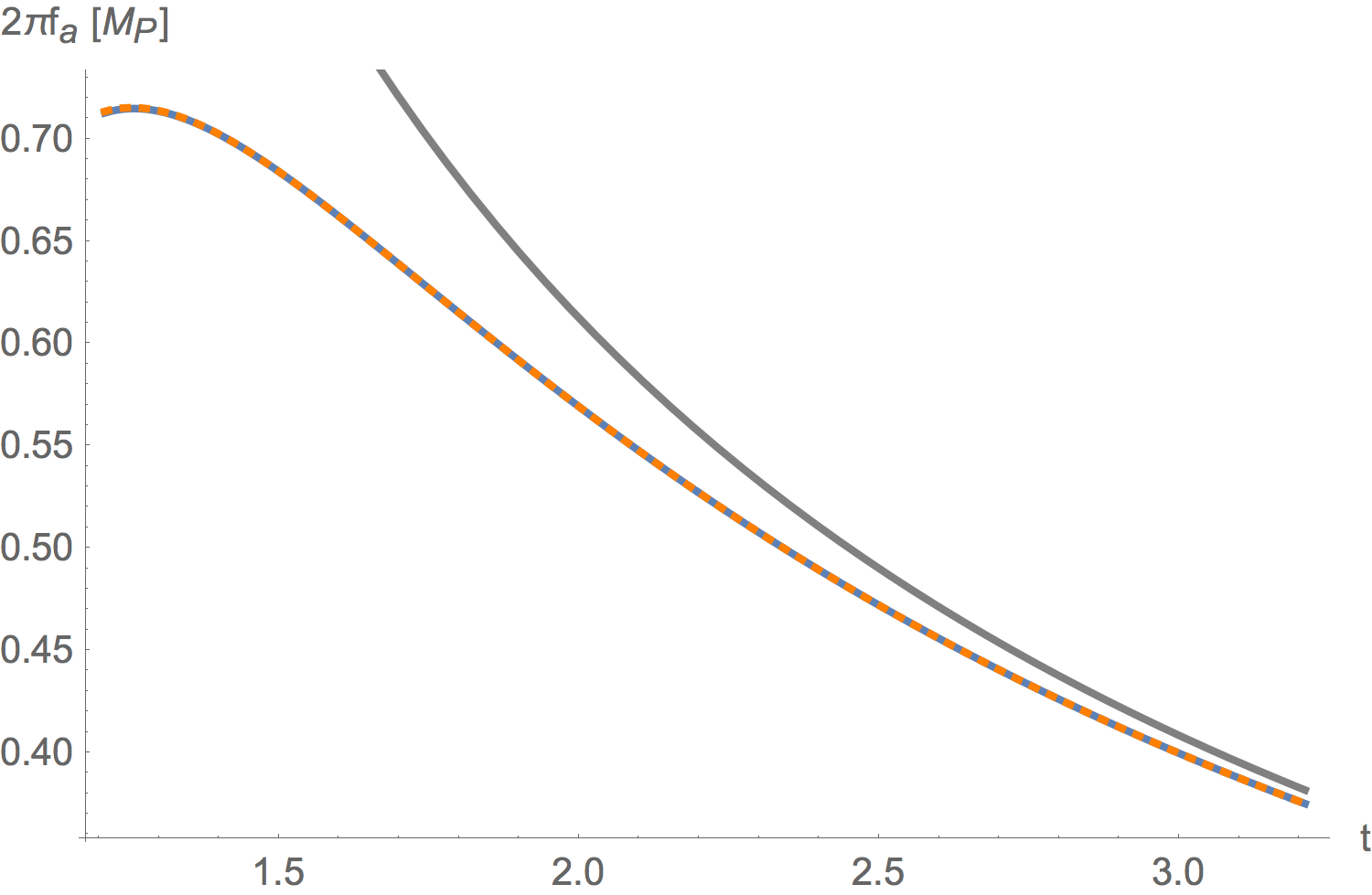}
\end{center}
\caption{Plot of the axion field range for the one-parameter quintic threefold without $\alpha'$-corrections (grey), for the full tree-level K\"ahler potential (blue), and including instanton contributions up to order ten (dotted orange). }
\label{fig:kaehlermetricquintic}
\end{figure}

The first correction term is the classical perturbative $\alpha'^{3}$-correction, which modifies the K\"ahler potential to
\begin{equation}
K=-\log{\left( \frac{20}{3} \left(-i\frac{(T - \bar{T})}{2} \right)^3+ \frac{50\zeta(3)}{\pi^3}\right)}\ .
\end{equation}
As shown in Figure~\ref{fig:kaehlermetricquintic}, this has the effect of
moderating the increase in the field range as $t$ reduces.
We also show in Figure~\ref{fig:kaehlermetricquintic} the effect of including all instanton corrections to give the
full non-perturbative K\"ahler metric. Surprisingly, these have minimal effect on the field
range for any given value of $\langle t \rangle$.\footnote{The reason is that the instanton corrections give sinusoidal corrections to the K\"ahler metric, so even when the correction to the metric itself is large the correction to the integrated field range is small.} However, the instanton corrections do have a profound effect on the maximum field range,
but this is through modifying the structure of the moduli space via the mirror map that describes the geometry of K\"ahler moduli space.

As described in detail in the classic paper~\cite{Candelas:1990rm}, mirror symmetry ensures that the moduli space of the K\"ahler modulus is equivalent to the moduli space of
the complex structure modulus on the mirror quintic. The $\mbb{Z}_5$ quotient symmetry ensures that the moduli space of the complex structure modulus
$U$ on the mirror quintic is the wedge of the complex plane given by ${\rm arg}(U) \le 2 \pi / 5$.
The mirror map (given explicitly as equation (5.9) in~\cite{Candelas:1990rm}) describes the map from $U \to T$ and shows how the fundamental
$U$ domain is mapped into K\"ahler moduli space.

\begin{figure}
\begin{center}
\includegraphics[width=0.4\textwidth]{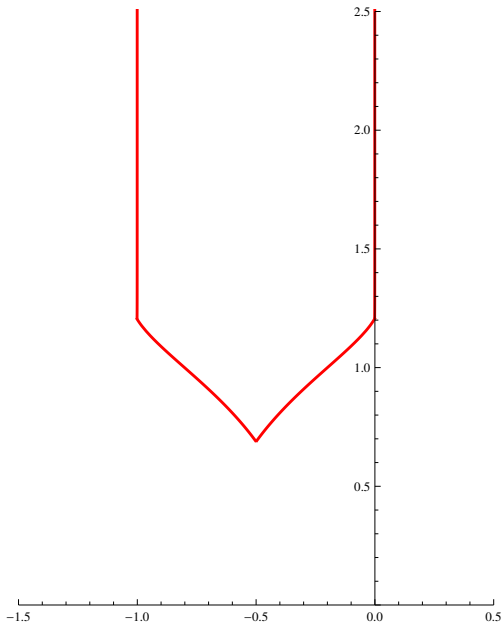}
\end{center}
\caption{Fundamental domain of the K\"ahler modulus of the quintic.}
\label{fig:quinticfundamentaldomain}
\end{figure}

The upshot is that the fundamental region of K\"ahler moduli space is, in a manner analogous to the dilaton-axion system, a rectangle coming down from infinite
volume but capped at small volume. This is illustrated in Figure~\ref{fig:quinticfundamentaldomain}. The same physics is seen; there is a maximal, approximately Planckian,
value that the axion field range attains within the fundamental domain (for the quintic this is 0.7 $M_P$). Beyond this, trajectories can be mapped back into non-axionic
trajectories within the fundamental domain.

\subsection{More One-Parameter Models}

Following the above example of the quintic, we also want to perform a similar analysis for other one-parameter Calabi-Yaus for which
mirror symmetry data on the mirror map is available. For this purpose we
use the Calabi-Yau manifolds described in first~\cite{Klemm:1992tx}, then~\cite{Klemm:1993jj}, and finally~\cite{Kawada:2015vwe,Braun:2015jdy}.

The first paper contains the following examples (including the quintic above)
\begin{eqnarray}
\nonumber M_{k=5}&=& \left\{ x_i\in\mathbb{P}_{[1,1,1,1,1]}\ |\ W_0=x_0^5+x_1^5+x_2^5+x_3^5+x_4^5=0\right\},\\
\nonumber M_{6}&=& \left\{ x_i\in\mathbb{P}_{[2,1,1,1,1]}\ |\ W_0=2x_0^3+x_1^6+x_2^6+x_3^6+x_4^6=0\right\},\\
\nonumber M_8&=& \left\{ x_i\in\mathbb{P}_{[4,1,1,1,1]}\ |\ W_0=4x_0^2+x_1^8+x_2^8+x_3^8+x_4^8=0\right\},\\
\nonumber M_{10}&=& \left\{ x_i\in\mathbb{P}_{[5,2,1,1,1]}\ |\ W_0=5x_0^2+2x_1^5+x_2^{10}+x_3^{10}+x_4^{10}=0\right\}.
\end{eqnarray}
For these examples the prepotential takes the form
\begin{equation}
F=-\frac{y}{6}T^3+\frac{1}{2}\kappa_2 T^2+\kappa_1 T+\kappa_0-\frac{1}{(2\pi i)^3} \sum_{k,d=1}^\infty n_k \frac{e^{2\pi i d k T}}{d^3}~,
\end{equation}
where the parameters for each model are given as follows:
\begin{equation}
\def\arraystretch{1.5}
\begin{array}{c|c|c|c}
y & \kappa_2 & \kappa_1 & \kappa_0 \\ \hline
 5 & -\frac{11}{2} & \frac{25}{12} & -\frac{25 i \zeta (3)}{\pi ^3} \\
 3 & -\frac{9}{2} & \frac{7}{4} & -\frac{51 i \zeta (3)}{2 \pi ^3} \\
 2 & -3 & \frac{11}{6} & -\frac{37 i \zeta (3)}{\pi ^3} \\
 1 & -\frac{1}{2} & \frac{17}{12} & -\frac{36 i \zeta (3)}{\pi ^3} \\
\end{array}
\end{equation}

In \cite{Klemm:1992tx} the mirror map and duality symmetries are described in detail, in particular showing how these
restrict the K\"ahler moduli space to a fundamental region analogous to that already described for the cases of $SL(2, \mbb{Z})$ and
the quintic. As the behaviour is simpler to that of the quintic, we do not show the precise plots (available in the journal version of \cite{Klemm:1992tx}).

We instead list results, describing the maximal axion field range, the value of the K\"ahler modulus when this is attained, and the classical
volume this corresponds to.
\begin{equation}
\def\arraystretch{1.2}
\begin{array}{c|c|c|c|c}
{\rm Manifold} & {\rm Euler~Number} & {\rm Field~Range} & {\rm Cycle~size}/(2 \pi \sqrt{\alpha'})^2 & {\rm Volume}/(2 \pi \sqrt{\alpha'})^6 \\ \hline
 \mathbb{P}_{[2,1,1,1,1]} & -204 & 0.6 M_P & 1.4 &  1.37 \\
 \mathbb{P}_{[4,1,1,1,1]} & -296 & 0.44 M_P & 1.65  &  1.51 \\
 \mathbb{P}_{[5,2,1,1,1]} & -288 & 0.35 M_P & 2.05 & 1.44 \\
\end{array}
\end{equation}
As for the case of the quintic, the maximal field range corresponds to the smallest value of $t$ such that the entire axionic path $a \to a + 1$ remains
inside the fundamental region.

We now also consider the following CY's with $h^{1,1}=1$ described in~\cite{Klemm:1993jj}.
\begin{table}
\centering
\def\arraystretch{1.2}
\begin{tabular}{c|c|c|c}
Label & Hypersurface & $\chi(X)$ & $y$ \\ \hline
1 & $X_{4,4}\subset \mathbb{P}^5_{[1,1,2,1,1,2]} $ & -144 & 4 \\
2 & $X_{6,6}\subset \mathbb{P}^5_{[1,2,3,1,2,3]} $ & -120 & 1 \\
3 & $X_{4,3}\subset \mathbb{P}^5_{[2,1,1,1,1,1]} $ & -156 & 6 \\
4 & $X_{6,2}\subset \mathbb{P}^5_{[3,1,1,1,1,1]} $ & -256 & 4 \\
5 & $X_{6,4}\subset \mathbb{P}^5_{[3,2,2,1,1,1]} $ & -156 & 2 \\
\end{tabular}
\caption{List of one parameter models in~\cite{Klemm:1993jj}.}
\end{table}
In this case, the instanton numbers are calculated in \cite{Klemm:1993jj}, but the details of the mirror map, duality groups and fundamental regions
are not presented.

In evaluating the maximal field range, we therefore assume a similar feature holds for these examples as held for the above cases: that the boundary of the
fundamental region can be found by determining when the instanton sum causes the K\"ahler metric to diverge. Doing this, we then find the following field ranges
for the models listed:
\begin{equation}
\def\arraystretch{1.2}
\begin{array}{c|c|c|c|c}
{\rm Manifold} & {\rm Euler~Number} & {\rm Field~Range} & {\rm Cycle~size}/(2 \pi \sqrt{\alpha'})^2 & {\rm Volume}/(2 \pi \sqrt{\alpha'})^6 \\ \hline
 \mathbb{P}_{[1,1,2,1,1,1]} & -144 & 0.76 M_P & 1.18 &  1.09 \\
  \mathbb{P}_{[1,2,3,1,2,3]} & -120 & 0.54 M_P & 1.79 &  0.96 \\
  \mathbb{P}_{[2,1,1,1,1,1]} & -156 & 0.84 M_P & 1.05 &  1.16 \\
  \mathbb{P}_{[3,1,1,1,1,1]} & -256 & 0.58 M_P & 1.27 &  1.37 \\
  \mathbb{P}_{[3,2,2,1,1,1]} & -156 & 0.57 M_P & 1.51 &  1.15 \\
\end{array}
\end{equation}

In these examples, the precise physics that determines the maximal axionic field range involves the rate of growth of the instanton sum, which determines the radius
of convergence of the large-$t$ expressions (and thus implicitly the fundamental region of moduli space). The $\alpha'^{3}$-corrections serves to moderate the growth in field
range as we approach small radius.

The final examples are based on results in~\cite{Braun:2009qy,Kawada:2015vwe,Braun:2015jdy}.
In comparison to the previous examples, these have only a small number of geometric moduli.

These manifolds are obtained by modding out freely acting symmetries in a CICY, leading to manifolds with $h^{1,1} = 1$ and $h^{2,1}=\{1,3,4,5\}$, and so $\chi(M)=\{0,-4,-6,-8\}$. The large complex structure expansion breaks down at the closest conifold point which can be identified from the Picard-Fuchs operator. Utilising the mirror map, we can identify the corresponding value for $T.$
\begin{equation}
\def\arraystretch{1.2}
\begin{array}{c|c|c|c|c|c}
{\rm Manifold} & \chi & y & {\rm Field~Range} & {\rm Cycle~size}/(2 \pi \sqrt{\alpha'})^2 & {\rm Vol.}/(2 \pi \sqrt{\alpha'})^6 \\ \hline
 (1,1) & 0 & 4 & 2.81 M_P & 0.45 &  0.06 \\
  (1,3) & -4 & 12 & 2.79 M_P & 0.44 &  0.17 \\
  (1,4) & -6 & 18 & 3.25 M_P & 0.33 &  0.11 \\
  (1,5) & -8 & 3 & 2.15 M_P & 0.57 &  0.09
\end{array}
\end{equation}

The combination of a small Euler number, a large triple intersection, and a slow growth in instanton number allows convergence down to rather small radii. We find the maximal field range in the (1,4) manifold at $t=0.33 (2 \pi \sqrt{\alpha'})^2$, equivalent to a classical
volume of $0.11 (2 \pi \sqrt{\alpha'})^6$, leading to a field range of $3.25 M_P$. Interestingly, we see that even in the case of vanishing Euler number, the field range is bounded due to the presence of a conifold locus.

\subsection{General Discussion}

For a general one-parameter model, we can calculate the perturbative K\"ahler potential starting from the prepotential
\begin{equation}
F=-\frac{1}{6}y T^3+\frac{1}{2}\kappa_{2} T^2+\kappa_1 T + \frac{\zeta(3)\chi}{2 (2\pi i)^3},
\end{equation}
leading to the perturbative K\"ahler potential
\begin{equation}
K=-\log{\left(\frac{4~y ~t^3}{3}-\frac{\chi\zeta{(3)}}{4\pi^3}\right)}.
\end{equation}
As $t$ decreases from infinity, the resulting field range is maximised at (assuming $\chi/y<0$)
\begin{equation}
t_{\rm max}=\left(-\frac{21+9\sqrt{5}}{32\pi^3}\frac{\chi\zeta{(3)}}{y}\right)^{1/3}~.
\end{equation}
Numerically the K\"ahler metric and field range takes the value
\begin{equation}
K_{T\bar{T}}(t_{\rm max})\approx 2.60 \left(-\frac{y}{\chi}\right)^{2/3}\ ,\qquad 2 \pi f_a(t_{\rm max})\approx 2.28 \left(-\frac{y}{\chi}\right)^{1/3} M_P ~.
\end{equation}
We see that these maximal values depend on two parameters, the Yukawa coupling $y$ and the Euler number of the Calabi-Yau $\chi$. In particular one can achieve situations where $t_{\rm max}<1$ and the associated value of the K\"ahler metric $K_{T\bar{T}}>1$ and subsequently the axion field range obeys $2 \pi f_a> M_P.$

This illustrates the reason why the final examples with only a few moduli produced the largest field ranges; the combination of relatively
large triple intersection numbers and relatively
small Euler numbers both work to increase the field range.

This set of examples is supportive of the notion that string theory really does cap axion field ranges at values around $M_P$ (and not, say, around $8 \pi M_P$).
The physics operating here is a set of duality transformation that restrict the fundamental region of moduli space such that axion decay constants
cannot grow in an unbounded fashion. Even for cases where the instanton sum is absent (as for the dilaton/axion system), it is the presence of these
duality relations that enforce the presence of a fundamental region which caps the field range.

\section{Two parameter models}
\label{sec:twomoduli}
Let us now discuss examples with two K\"ahler moduli for which the prepotential is known in the literature. Before examining the individual examples, we would like to highlight the differences to the one parameter case. Similarly, we can start with a general prepotential:
\begin{equation}
F=-\frac{1}{6}y_{ijk}T_i T_j T_k+\frac{1}{2}\kappa_{ij}T_i T_j+\kappa_i T_i+\frac{\zeta(3)\chi}{2 (2\pi i)^3}+\sum_\beta n_\beta\ {\rm Li}_3(q^\beta)~.
\end{equation}
From the prepotential we can calculate the K\"ahler potential and metric in general. In the case of vanishing instanton contributions we can write the K\"ahler potential as
\begin{equation}
K=-\log{\left(\frac{4}{3}\left(y_{11}t_1^3+y_{12}t_1^2t_2+y_{21}t_1t_2^2+y_{22}t_2^3\right)-\frac{\chi\zeta{(3)}}{4\pi^3}\right)}.
\label{eq:kaehlerpotentialtwo}
\end{equation}
How do such a non-trivial K\"ahler potential and metric affect the axion decay constants? To see this, we need to diagonalise the kinetic terms. This leads to the following type of kinetic terms
\begin{equation}
{\cal L}_{\rm kin}=\sum_i\lambda_i \partial_\mu \tilde{T}_i\partial^\mu \bar{\tilde{T}}_i.
\end{equation}
where $\lambda_i$ denote the eigenvalues of the K\"ahler metric.

The notion of an axionic field range is not uniquely defined for multi-axion examples (there are several good bases for the axion lattice, and including instanton effects the
field range for one axion depends on the vev of the other axions).
As an illustrative measure, we shall define the decay constants to be given by these eigenvalues of the K\"ahler metric
\begin{equation}
2\pi f_i=\sqrt{2\lambda_i}M_P ~.
\label{eq:decayconstants2}
\end{equation}
As in the one modulus case, the formula for the decay constants~\eqref{eq:decayconstants2} is modified when including the instanton contributions to
\begin{equation}
2\pi f_i=\int_{a_1,a_2=0}^1\sqrt{2\lambda_i}M_P~ da_i~.
\end{equation}
A more detailed motivation for this choice for the decay constants can be found in Appendix~\ref{app:decayconstants}.

As the general eigenvalues are lengthy expressions without a clear structure, we present some examples that have appeared in the literature and analyse the decay constants that appear in this context. We shall discuss three examples in turn $\mathbb{P}^4_{(1,1,1,6,9)}[18],$ $\mathbb{P}^4_{(1,1,2,2,2)}[8],$ and $\mathbb{P}^4_{(1,1,2,2,6)}[12].$ All of these examples are hypersurfaces in weighted projective spaces.

\subsection{$\mathbb{P}^4_{(1,1,1,6,9)}[18]$}
One of the canonical two moduli examples is the weighted projective space $\mathbb{P}^4_{(1,1,1,6,9)}[18]$ whose moduli space was presented and discussed in~\cite{Candelas:1993dm}.  We are interested in type IIA string theory on this manifold with Hodge numbers $(2,272)$ which is described by the following degree 18 polynomial
\be
p=x_1^{18}+x_2^{18}+x_3^{18}+x_4^3+x_5^2-18\psi x_1x_2x_3x_4x_5-3\phi x_1^6x_2^6x_3^6\ .
\ee
The prepotential is given by
\begin{equation}
F=-\frac{1}{6}\left(9T_1^3+9T_1^2T_2+3T_1T_2^2\right)+\frac{1}{4}\left(9T_1^2+3T_1T_2\right)+\frac{17}{4}T_1+\frac{3}{2}T_2
-\frac{135i\zeta{(3)}}{4\pi^3}+\sum_{i,j}n_{ij}\ {\rm Li}_3\left(q_1^iq_2^j\right),
\label{eq:p11169prepotential}
\end{equation}
where the first numbers associated to the instanton contributions can be found on page~44 of~\cite{Candelas:1993dm} and in Appendix~\ref{app:perioddetails}.

Figure~\ref{fig:p11169ev} shows the eigenvalues of this K\"ahler metric. From the plot we get the impression that it is possible to have large trans-Planckian decay constants in some regions of parameter space. However, are these regions which are actually accessible?
\begin{figure}
\begin{center}\includegraphics[width=0.48\textwidth]{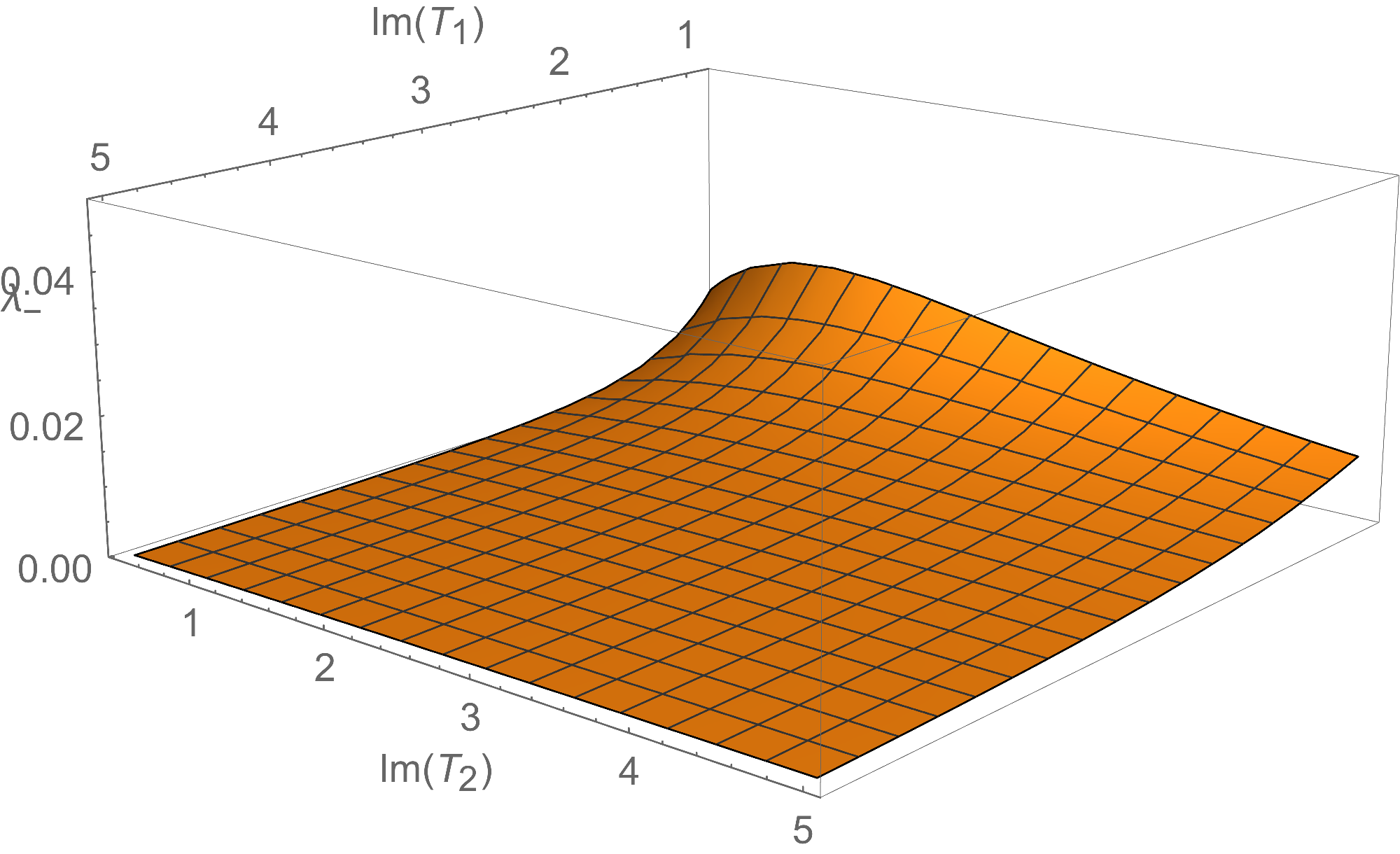}\includegraphics[width=0.48\textwidth]{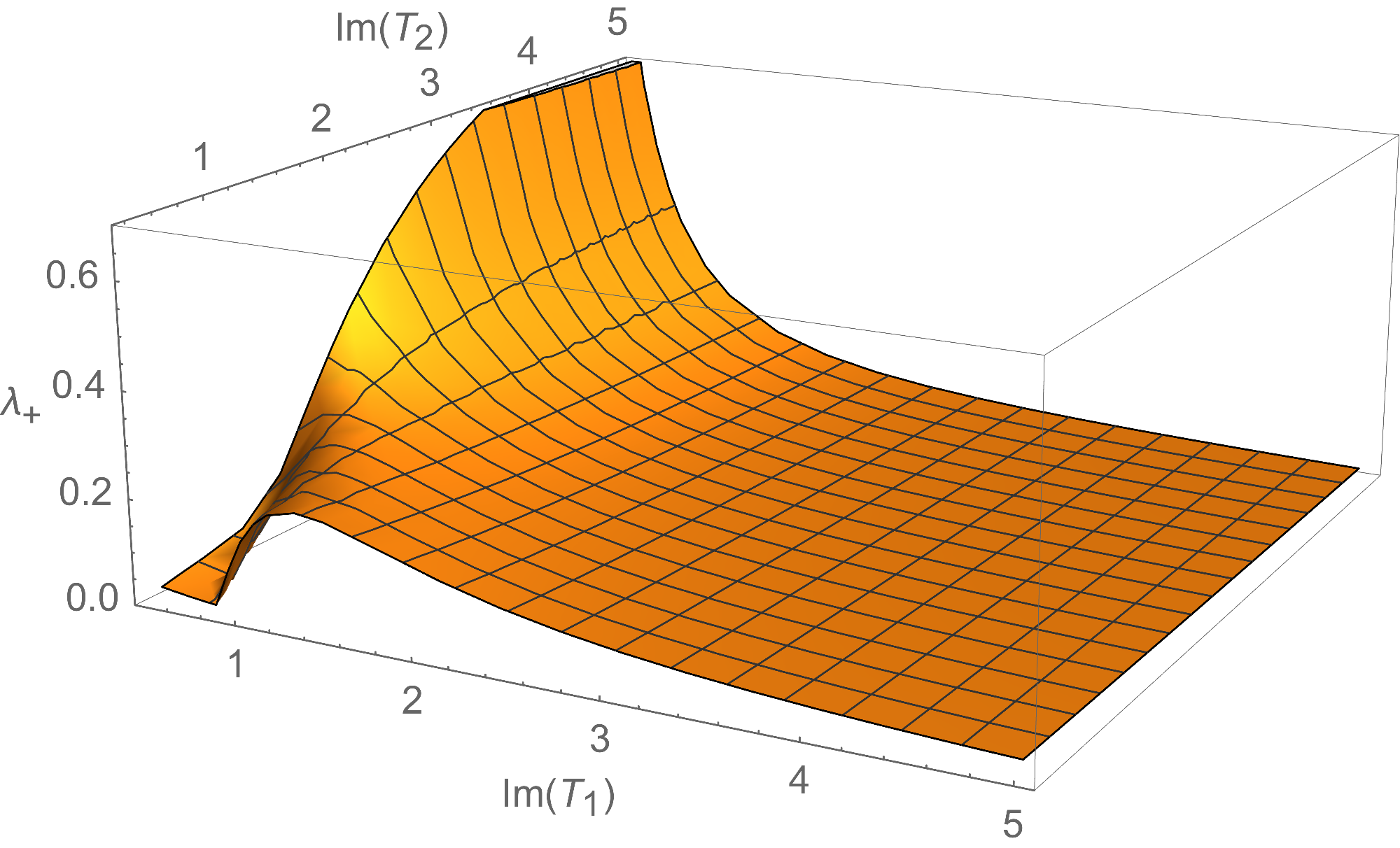}\end{center}
\caption{Eigenvalues of the K\"ahler metric for $\mathbb{P}^4_{(1,1,1,6,9)}[18]$ derived from the prepotential in~\eqref{eq:p11169prepotential} with no instanton contributions.}
\label{fig:p11169ev}
\end{figure}

In the case of the quintic one-parameter threefold, the large complex structure expansion is valid up to the conifold point which is located at $\psi=1.$ Here, there are similar boundaries located at special points in moduli space where the large complex structure expansion of the periods is no longer valid. In the case of $\mathbb{P}_{(1,1,1,6,9)}^4$ they are located at
\begin{equation}
(\rho^6+\phi)^3=1\ , \qquad \phi^3=1\ ,
\end{equation}
where $\rho=(3^4 2)^{1/3}\psi.$
As for the quintic, we can now find the corresponding values for $T_{1,2}$ by utilising the mirror map and establish the boundary locus in this parametrisation. The mirror map was provided in~\cite{Candelas:1993dm} and is summarised in Appendix~\ref{app:perioddetails} for completeness. Let us discuss these two loci in turn.

\begin{figure}
\includegraphics[width=0.45\textwidth]{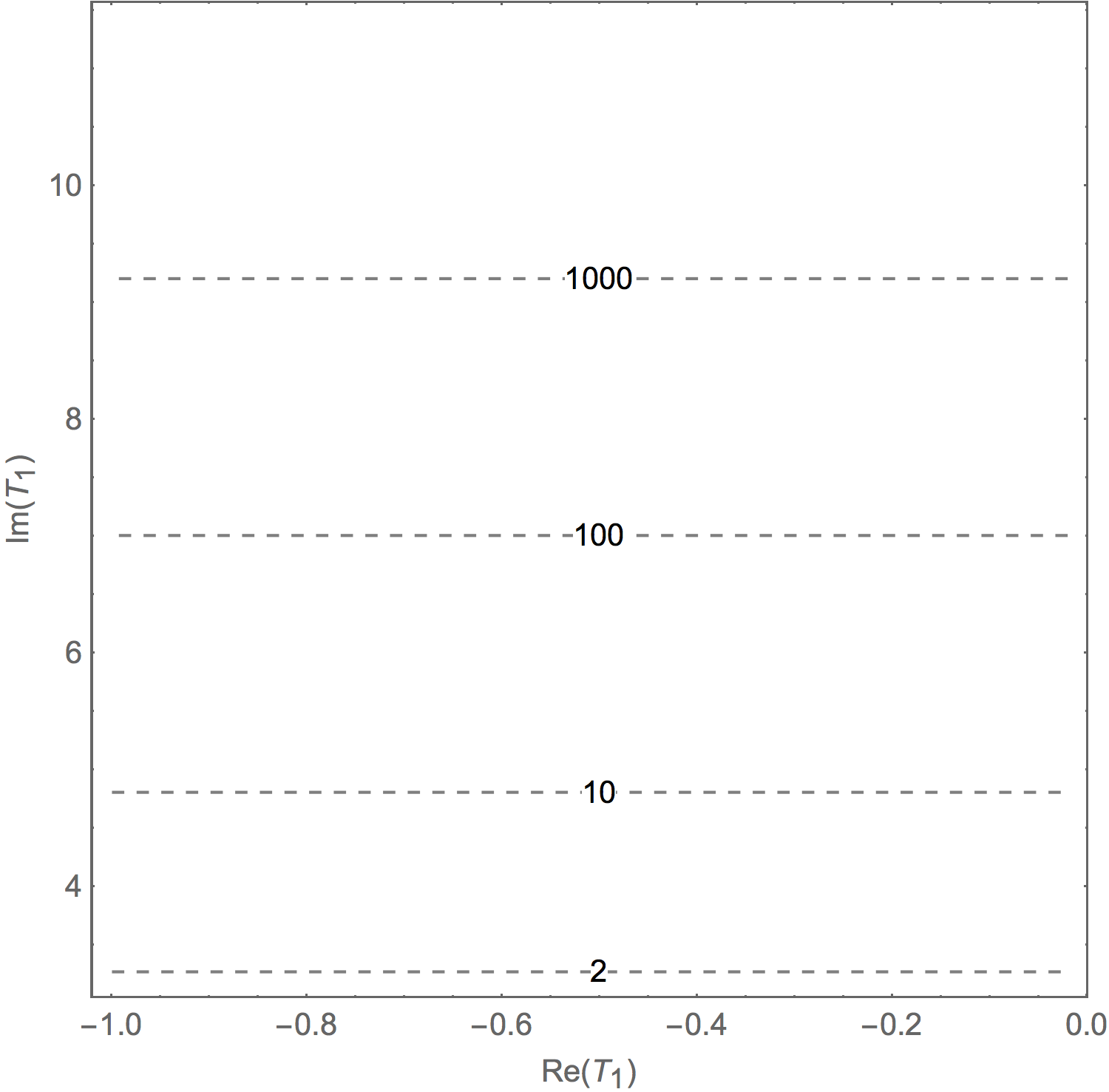}
\includegraphics[width=0.45\textwidth]{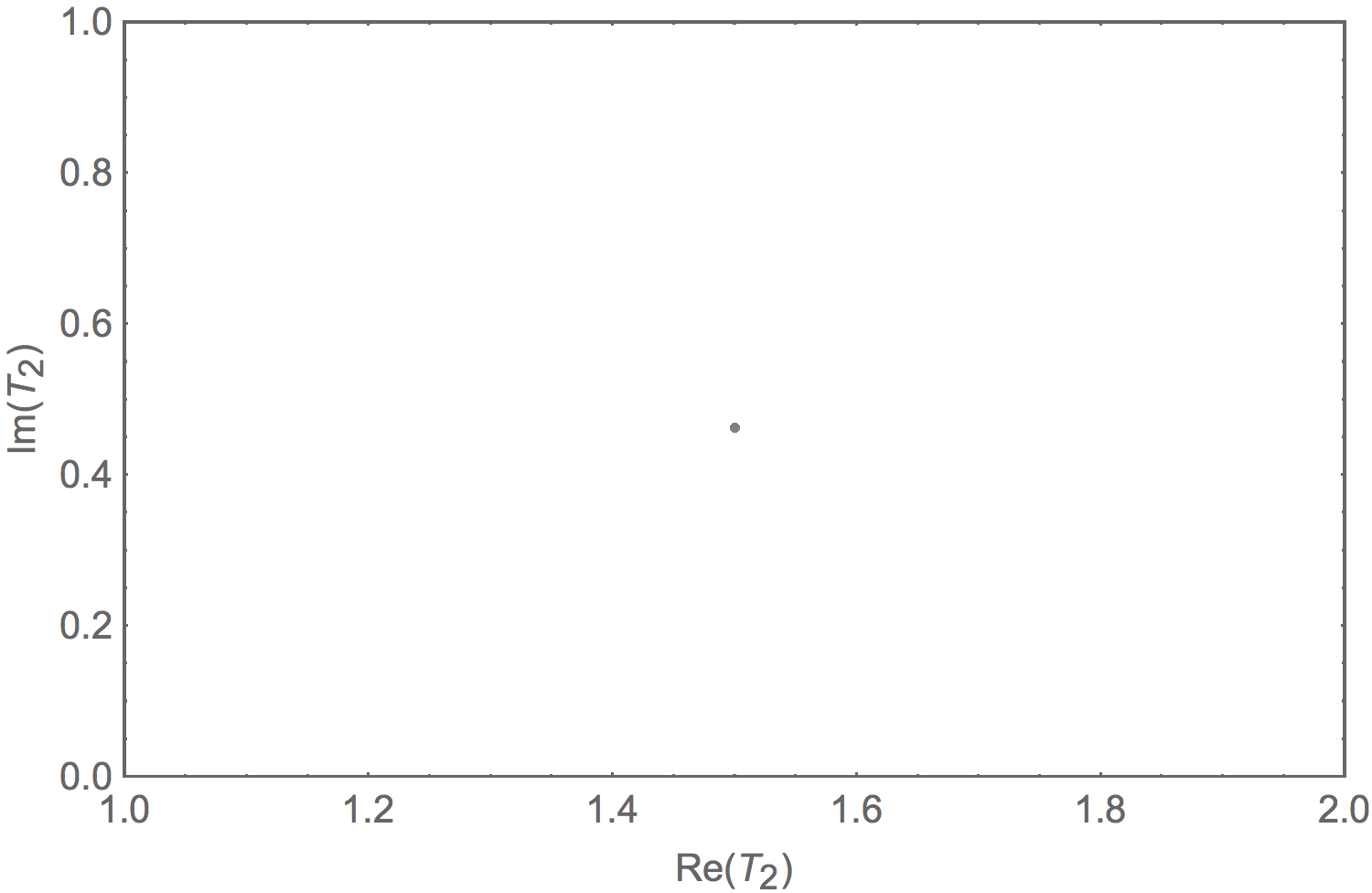}
\caption{Contour lines of $|\psi|={\rm const.}$ for $T_{1}$ (left) and $T_{2}$ (right). $T_2$ is fixed and $T_1$ varies by varying $\psi.$ For small $\psi$ (not shown here) the behaviour becomes non-trivial. There are in principle three distinct solutions to $\phi^3=1$, however the contours remain the same for sufficiently large values for $|\psi|.$}
\label{fig:modspace1}
\end{figure}

\begin{figure}
\includegraphics[width=0.45\textwidth]{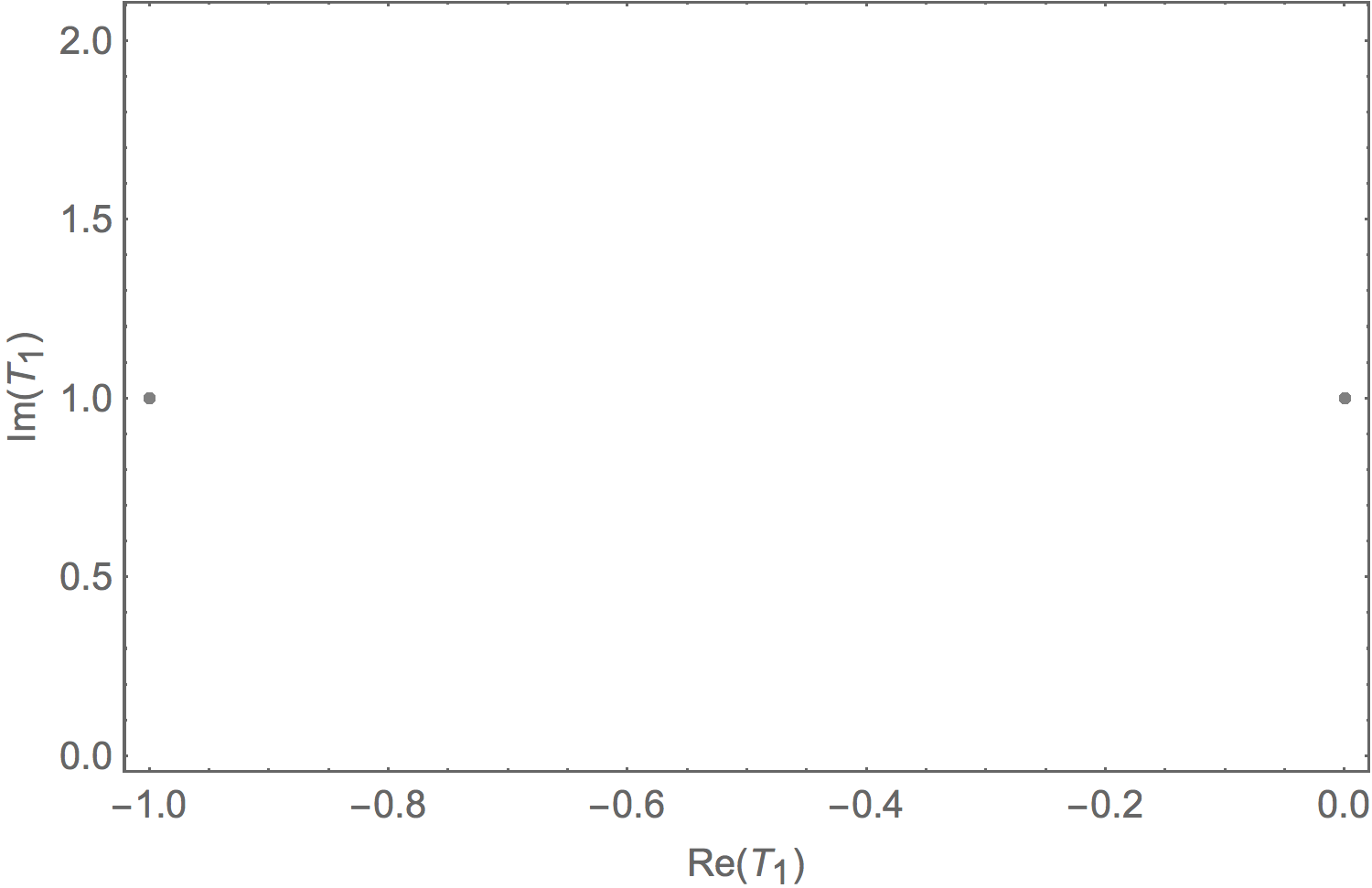}
\includegraphics[width=0.45\textwidth]{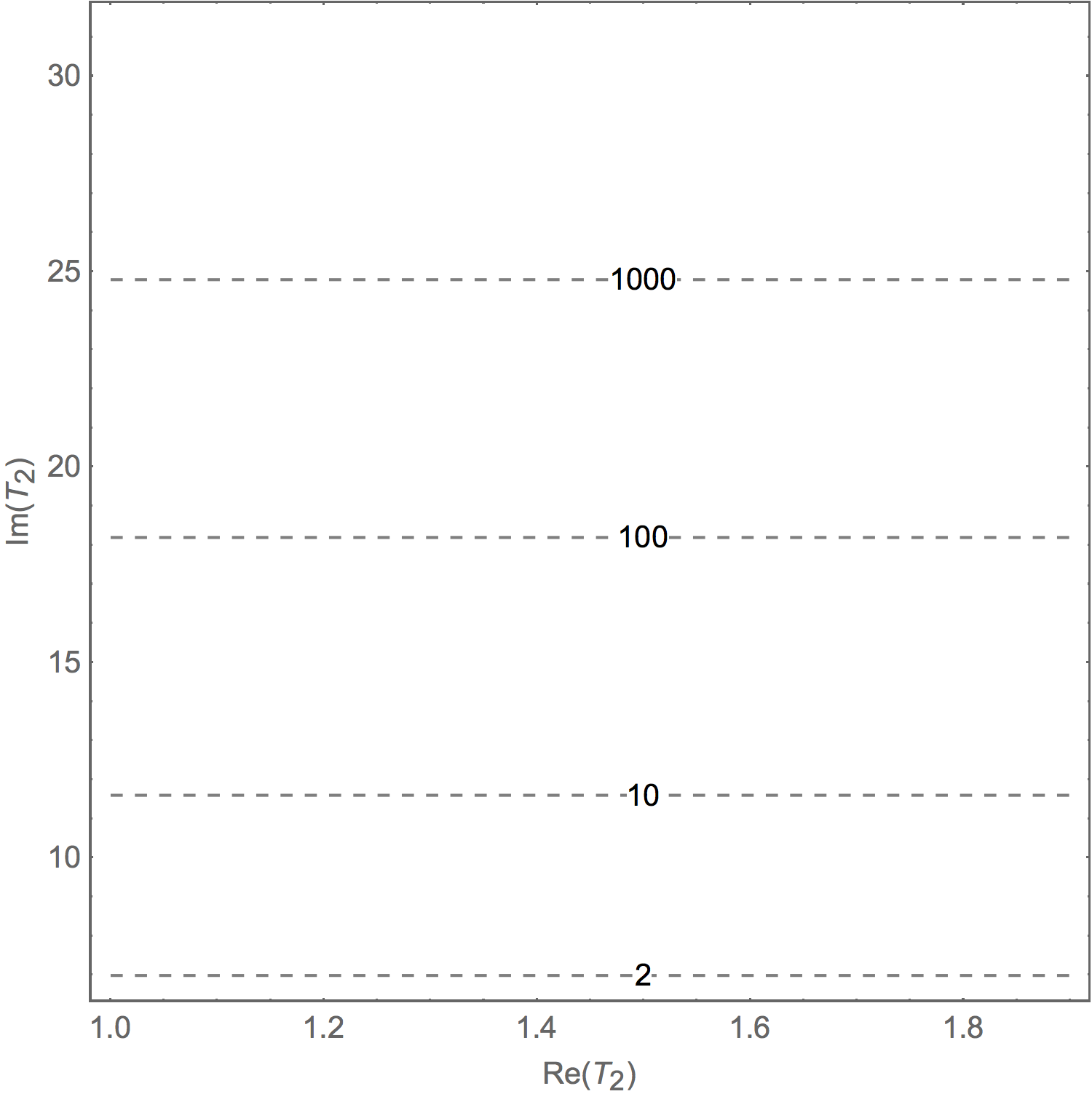}
\caption{Contour lines of $|\psi|={\rm const.}$  for $T_{1}$ (left) and $T_{2}$ (right). $T_1$ is fixed and $T_2$ varies by varying $\phi$ or $\psi$ respectively.}
\label{fig:modspace2}
\end{figure}

For the hypersurface $\phi^3=1,$ a sketch of the parameter space is shown in Figure~\ref{fig:modspace1}. It shows that by decreasing $T_2$ towards the critical value $T_2\approx 1.50+0.46 i $ we hit this special point in moduli space. As in the one modulus case we cannot decrease the imaginary part of $T_2$ below this point with this expansion of the periods. The large complex structure expansion breaks down.
For the locus $(\rho^6+\phi)^3=1,$ the contour lines are shown in Figure~\ref{fig:modspace2} where we find a similar behaviour as for the locus $\phi^3=1$ with the difference that $T_1$ remains almost constant (slightly decreasing with increasing $|\psi|$) at around $T\approx i$ whereas $T_2$ is free to vary.

In analogy to the effect in the one-modulus case, the instanton contributions do not change the results significantly.
However, they do set the boundary of the large complex structure expansion, which then restricts the decay constants significantly at small values for $T_i.$

Both conifold loci bound this moduli space at small values for $T_i.$ In particular the regions with large decay constants at large values for $t_2$ and small values for $t_1$ are excluded. The maximal value of the decay constants (corresponding to the larger eigenvalue) is found to be
\begin{equation}
2\pi f_{\rm max}=0.71 M_P~,
\end{equation}
where ${\rm Im}(T_1)=1.00$ and ${\rm Im}(T_2)=4.24.$

\subsection{Further examples}
The moduli space of the following two examples was presented and discussed in~\cite{Candelas:1993dm}. The first Calabi-Yau manifold, with Hodge numbers $(2,86),$ is given as the following degree eight polynomial in the weighted projective space $\mathbb{P}^4_{(1,1,2,2,2)}$
\begin{equation}
W=x_1^8+x_2^8+x_3^4+x_4^4+x_5^4-8\psi x_1x_2x_3x_4x_5 -2 \phi x_1^4 x_2^4~.
\end{equation}
The non-vanishing Yukawa couplings are given by
\begin{equation}
y_{11}=8~,\qquad y_{12}=12~.
\end{equation}
The first instanton contributions are summarised in Appendix~\ref{app:perioddetails}. The loci where the large complex structure expansion breaks down are given by
\begin{equation}
\phi^2=1~,\qquad 8 \psi^4+\phi=\pm 1~.
\end{equation}
The relevant details from the mirror map from~\cite{Candelas:1993dm} are summarised in Appendix~\ref{app:perioddetails}.

The second example is a Calabi-Yau with Hodge numbers $(2,128)$ which is given as the hypersurface in the weighted projective space $\mathbb{P}^4_{(1,1,2,2,6)}$ with degree 12
\begin{equation}
W=x_1^{12}+x_2^{12}+x_3^6+x_4^6+x_5^6-12\psi x_1x_2x_3x_4x_5-2\phi x_1^6x_2^6
\end{equation}
The non-vanishing Yukawa couplings are given by
\begin{equation}
y_{11}=4~,\qquad y_{12}=6~.
\end{equation}
The first instanton coefficients $n_{ij}$ are given in Appendix~\ref{app:perioddetails}.
The interesting points are similar and correspond to $\phi^2=1,$ $864 \psi^6+\phi=\pm 1.$ Details from the mirror map from~\cite{Candelas:1993dm} are summarised in Appendix~\ref{app:perioddetails}.\footnote{Some useful details can also be found in~\cite{Giryavets:2003vd}.}

For both examples we run the analogous analysis as for $\mathbb{P}^4_{1,1,1,6,9}[18],$ the corresponding Figures of the eigenvalues of the K\"ahler metric are similar. We find the following maximal decay constants
\begin{equation}
\def\arraystretch{1.3}
\begin{array}{c| c| c | c}
& f_{\rm max} & {\rm Im}(T_1)& {\rm Im}(T_2) \\ \hline
\mathbb{P}_{11222}^4[8] & 1.4 M_P & 0.71 & \infty \\ \hline
\mathbb{P}_{11226}^4[12] & 0.99 M_P & 1.015 & \infty
\end{array}
\end{equation}
We note that also large values for ${\rm Im}(T_2)$ reproduce similar field ranges as the asymptotic value.

We have seen that the general K\"ahler potential~\eqref{eq:kaehlerpotentialtwo} can lead to large eigenvalues (i.e.~significantly trans-Planckian decay constants) before taking into account constraints on the moduli space. These constraints, arising from the breakdown of the large complex structure expansion (e.g.~due to conifold loci in the CY moduli space), bound the decay constants to be comparable to $M_P.$ Again the instanton sum is found to have a negligible effect in our examples.

\section{Connection to (field theory) models of inflation}
\label{sec:connection}
In this article we have focused on the axion field range arising from the K\"ahler potential and the associated eigenvalues of the K\"ahler metric without discussing the scalar potential. For clarity, let us connect this discussion to field theory models of inflation which would precisely exhibit this behaviour.

Throughout this section we make the simplifying assumption that the K\"ahler potential which has been obtained in a context of ${\cal N}=2$ compactifications survives as the K\"ahler potential upon breaking to ${\cal N}=1$ SUSY. In an ${\cal N}=1$ framework we can discuss how such a non-trivial K\"ahler potential can lead to known models of axion inflation.

Taking the one modulus case with a kinetic term as discussed in Section~\ref{sec:onemodulus}, a potential of the type
\begin{equation}
V=\Lambda^4\left(1-\cos{(\phi)}\right)
\end{equation}
leads to a setup with precisely the decay constants as discussed. The two axion setup can be brought into the following form by appropriate field re-definitions
\begin{equation}
{\cal L}=K_{ij}(\partial \phi_i)(\partial \phi_j)-\Lambda^4\sum_{i=1}^2 \left(1-\cos{(\phi_i)}\right).
\label{eq:lagrangiantwoaxion}
\end{equation}
In this case the eigenvalues of the non-trivial K\"ahler potential set the axion decay constants. The effect of axion decay constant alignment~\cite{Kim:2004rp, Kappl:2014lra} precisely occurs if one of the associated eigenvalues of the kinetic mixing matrix becomes large (see~\cite{Burgess:2014oma} for a detailed discussion). In our analysis of the two moduli case we precisely capture the possibility of this effect. Note that from~\eqref{eq:lagrangiantwoaxion} there are other possibilities to generate decay constant alignment by generating a non-trivial axion potential. This is always an additional possibility which is not captured in our discussion.

However, the two potentials described above can be obtained by generating a potential from (simple) non-perturbative effects (e.g.~pure gaugino condensation), for instance with a superpotential of the following type
\begin{equation}
W=W_0+\sum_i A_i e^{-a_i T}~ .
\end{equation}
In particular, in the case of two moduli it is not required to arrange for very particular type of non-perturbative effects to reach (trans-)Planckian decay constants.\footnote{In general, there are additional mechanisms which can lead to decay constant alignment by considering more sophisticated brane setups with large wrapping number and appropriate units of flux turned on (see for instance~\cite{Long:2014dta}). Also, very special instanton configurations can lead to alignment in field theory~\cite{Choi:2014rja}. This could in principle generate alignment via the superpotential.} In other words, simple, low-rank gaugino condensates would give rise to a setup with large field inflation if the eigenvalue structure of the K\"ahler metric is appropriate.

Along similar lines one can envisage to compare with models that involve large number of axion fields (e.g.~\cite{Dimopoulos:2005ac,Long:2014fba}). However, the analysis of Calabi-Yau manifolds with such a large number of complex structure moduli is beyond the scope of this paper.

\section{Conclusion}
\label{sec:conclusion}
If there is a bound on axion field ranges in string theory, it will also hold in regions of strong coupling or small radii.
 As presented in this article it is possible to use mirror symmetry results to make a quantitative study of field ranges
 in the context of ${\cal N}=2$ Calabi-Yau compactifications of type II string theory, even in such strong coupling regions.

The largest field ranges we have found so far is $3.25~M_P$ (one modulus CY) and $1.4~M_P$ (two moduli CY), and it is worth noting that our results
do suggest that it is around $M_P$ (and not for example $16 \pi M_P$) that the critical value for field ranges lie.
However our exploration of Calabi-Yau geometries is very limited at this stage. It would be very intriguing to analyse large classes of the (available) Calabi-Yau manifolds to test for the possibility of large axion decay constants. The steps for this analysis are clear and involve the information that was used in the previous sections: the large complex structure prepotential (in particular the tree-level Yukawa couplings) and the singularity structure (i.e.~to know the range of validity of the large complex structure regime). As the tree-level Yukawa couplings are related to triple intersection numbers on the respective mirror manifold, it is clear that any upper bound on intersection numbers would influence the bound on the possible axion decay constants. Together with a general argument on the appearance of singularities in moduli space that shield off regions corresponding to large decay constants, a `bound' along the lines of our analysis can be established.

 We have found that the higher order instanton corrections have a negligible effect in our examples which is in accordance of the recent work~\cite{Kappl:2015esy,Choi:2015aem}. The effect that is capping the field range are duality symmetries which are of clear string theory origin.

It would also be interesting to see how this approach can be applied in the context of F-theory where open string moduli become part of complex structure moduli and to which extent the results on the possible field range can be extended to situations with open string moduli as well. We hope to return to some of these questions in the future.

\section*{Acknowledgments}
We would like to thank Philip Candelas and Markus Rummel for valuable discussions. This project is funded in part by the European Research Council as Starting Grant 307605-SUSYBREAKING. JC is also funded by a Royal Society University Research Fellowship. This work was performed in parts at the Aspen Center for Physics, which is supported by National Science Foundation grant PHY-1066293.

\appendix

\section{Conventions}
\label{sec:appconventions}
In this Appendix we summarise our conventions, in particular for the periods and the associated metric around the large complex structure limit.

The underlying K\"ahler potential for the IIB complex structure moduli (IIA K\"ahler moduli) is given by
\begin{equation}
K=-\log{\left(-i~\Pi^\dagger.\Sigma.\Pi\right)},
\end{equation}
where
\begin{equation}
\Sigma=\left(
\begin{array}{c c}
0& \bf{1}\\
-\bf{1} & 0
\end{array}
\right).
\end{equation}
The period vector $\Pi$ around the large complex structure limit is given by
\begin{equation}
\Pi=\left(\begin{array}{c}
1\\
T^i\\
2F-T^i\partial_iF \\
F_i
\end{array}
\right),
\end{equation}
where $F$ is the prepotential. Around the large complex structure point, it can be written in the following form
\begin{equation}
F=-\frac{1}{6}y_{ijk}T^i T^j T^k+\frac{1}{2}\kappa_{ij}T^i T^j+\kappa_i T^i+\frac{\zeta(3)\chi}{2 (2\pi i)^3}+\sum_\beta n_\beta {\rm Li}_3(q^\beta)\ ,
\end{equation}
where $q^j=e^{2\pi i T^j}$ and the $\beta$ denotes an appropriate multi-index.
We are interested in the K\"ahler metric
\begin{equation}
K_{i\bar{j}}=\frac{\partial^2 K}{\partial T^i \partial \bar{T}^{\bar{j}}}\, .
\end{equation}
The large complex structure limit is defined by the existence of an exact shift symmetry for every complex structure modulus $T_i=a_i+i t_i$
\begin{equation}
T_i\to T_i+1\ .
\end{equation}

\section{Period details}
\label{app:perioddetails}
Here we summarise the technical details of the two moduli geometries discussed in the main text. All of them have appeared beforehand and are only listed for completeness.

\subsection{$\mathbb{P}^4_{(1,1,1,6,9)}[18]$}
The first instanton numbers are given by
\begin{equation}
\begin{array}{l|rrrrrr}
j\setminus k & 0 & 1 & 2 & 3 & 4 & 5\\ \hline
0 & & 3 & -6 & 27 & -192 & 1695 \\
1 & 540 & -1080 & 2700 & -17280 & 154440 &  \\
2 & 540 & 143370 & -574560 & 5051970 &  &  \\
3 & 540 & 204071184 & 74810520 &  &  &  \\
4 & 540 & 21772947555 &  & &  &  \\
5 & 540 &  &  &  &  &  \\
\end{array}
\end{equation}
In the large complex structure regime, the mirror map is given by
\begin{eqnarray}
\nonumber 2\pi i T^1&=& -\pi i-\log{\left(\frac{(18\psi)^6}{3\phi}\right)}+\frac{1}{\varpi_0}\sum_{k=0}^\infty \frac{(-3)^k(6k)!}{k! (2k)!(3k)!(18\psi)^{6k}}\left[A_kU_k(\phi)+Y_k(\phi)+N_k(\phi)\right]\\
2\pi i T^2&=& -3\pi i-\log{\left(3\phi\right)^3}-\frac{3}{\varpi_0}\sum_{k=0}^\infty \frac{(-3)^k(6k)!}{k! (2k)!(3k)!(18\psi)^{6k}}N_k(\phi)
\end{eqnarray}
where we used the following auxiliary functions
\begin{eqnarray}
\varpi_0&=&\sum_{k=0}^\infty\frac{(6k)!(-3)^k}{k!(2k)!(3k)!(18\psi)^{6k}}U_k(\phi)\\
\nonumber A_k&=& 6 \Psi(6k+1)-3\Psi(3k+1)-2\Psi(2k+1)-\Psi(k+1)\\
\nonumber U_k(\phi)&=&\phi^k\sum_{n=0}^{\left[k/3\right]}\frac{(-1)^nk!}{(n!)^3\Gamma(k-3n+1)(3\phi)^{3n}}\\
\nonumber Y_k(\phi)&=&\phi^k k!\sum_{n=0}^{\left[k/3\right]}\frac{(-1)^n}{(n!)^3(k-3n)!(3\phi)^{3n}}\left(\Psi(1+k)-\Psi(1+n)\right)\\
\nonumber N_k(\phi)&=&\phi^k k!\left(\sum_{n=0}^{\left[k/3\right]}\frac{(-1)^n\left[\Psi(n+1)-\Psi(k-3n+1)\right]}{(n!)^3(k-3n)!(3\phi)^{3n}}+\sum_{\left[k/3\right]+1}^\infty \frac{(-1)^{k+1}(3n-k-1)!}{(n!)^3(3\phi)^{3n}}\right)
\end{eqnarray}
\subsection{$\mathbb{P}^4_{(1,1,2,2,2)}[8]$}
The instanton numbers are given by
\begin{equation}
\begin{array}{l| rrrr}
j\setminus k & 0 & 1 & 2 & 3\\ \hline
1 & 0 & 0 & 0 & 0 \\
2 &  0 & 0 & 0 & 0 \\
3 & -1280 & 2560 & 2560 & -1280 \\
4&  -317864 & 1047280 & 15948240 & 1047280 \\
5 & -36571904 & 224877056 & 12229001216 & 12229001216 \\
6 & -3478899872 & 36389051520 & 4954131766464 & 13714937870784 \\
\end{array}
\end{equation}
The mirror map is given by

\begin{eqnarray}
\nonumber 2i\pi T_1&=&i\pi-\log{Z_1}-\sum_{n=1}^\infty\frac{(2n-1)!}{(n!)^2}Z_2^{-n}\\
\nonumber && +\frac{1}{2\varpi_0}\sum_{n=1}^\infty\frac{(4n)!(-1)^n}{(n!)^4}Z_1^{-n}[2A_n\hat{u}_n(Z_2)+2\hat{h}_n(Z_2)+\hat{f}_n(Z_2)]~,\\
2i\pi T_2&=&-\log{Z_2}+2\sum_{n=1}^\infty\frac{(2n-1)!}{(n!)^2}Z_2^{-n}-\frac{1}{\varpi_0}\sum_{n=1}^\infty \frac{(4n)!(-1)^n}{(n!)^4}Z_1^{-n}\hat{f}_n(Z_2)~,
\end{eqnarray}
where
\begin{eqnarray}
\nonumber &&Z_1=\frac{(8\psi)^4}{2\phi}~,\qquad Z_2=(2\phi)^2~,\\
\nonumber &&A_n=4[\Psi(4n+1)-\Psi(n+1)]~,\\
\nonumber &&\varpi_0(\phi,\psi)=\sum_{n=0}^\infty\frac{(4n)!(-1)^n}{(n!)^4(8\psi)^{4n}}u_n(\phi)~,\\
\nonumber &&\hat{u}_n=\frac{u_n}{(2\phi)^n}~,\\
\nonumber &&\hat{h}_n=\frac{h_n}{2(2\phi)^n}~,\\
&&\hat{f}_n=-\frac{\sqrt{\phi^2-1}f_n}{(2\phi)^n}~,
\end{eqnarray}
where the expansion for the fundamental period is valid for $|(\phi\pm 1)/(8\psi^4)|<1.$
The remaining auxiliary functions can be calculated using the following recursion relations:
\begin{eqnarray}
\nonumber &&n u_n=2(2n-1)\phi u_{n-1}-4(n-1)(\phi^2-1)u_{n-2}\ , \qquad u_0=1, t_1=2\phi\\
\nonumber &&n f_n=2(2n-1)\phi f_{n-1}-4(n-1)(\phi^2-1)f_{n-2}\ , \qquad f_0=0, f_1=4\\
\nonumber &&n g_n=2(2n-1)\phi g_{n-1}-4(n-1)(\phi^2-1)g_{n-2}\\
\nonumber &&\qquad\qquad -2u_n+8\phi u_{n-1}-8(\phi^2-1)u_{n-2}\ , \qquad g_0=0 , g_1=0\\
&&h_n=\phi f_n+g_n~.
\label{eq:auxiliaryfct}
\end{eqnarray}

\subsection{$\mathbb{P}^4_{(1,1,2,2,6)}[12]$}
The instanton numbers are
\begin{equation}
\begin{array}{l| rrr}
j\setminus k & 0 & 1 & 2\\ \hline
0 & 0 & 2 & 0 \\
1&  2496 & 2496 & 0 \\
2& 223752 & 1941264 & 223752 \\
3&  38637504 & 1327392512 & 1327392512 \\
4 &  9100224984 & 861202986072 & 2859010142112 \\
\end{array}
\end{equation}

The fundamental period in the large complex structure regime is given by
\begin{eqnarray}
\varpi_0&=&\sum_{n=0}^\infty \frac{(6n)!(-1)^n}{(n!)^3(3n)! (12\psi)^{6n}}u_n(\phi)~,~\left|\frac{\phi\pm1}{864\psi^6}\right|<1
\end{eqnarray}
where $u_n$ is the same function as above in~\eqref{eq:auxiliaryfct}.

In the regime where the large complex structure expansions of the periods is valid, the mirror map is given by
\begin{eqnarray}
\nonumber 2i\pi T^1&=&i\pi - \log{Y_1}\\
\nonumber &&-\sum_{n=1}^{\infty}\frac{(2n-1)!}{(n!)^2}Y_2^{-n}+\frac{1}{2\varpi_0}\sum_{n=1}^\infty\frac{(6n)!(-1)^n}{(n!)^3(3n)!}Y^{-n}_1\left(2B_n\hat{u}_n(Y_2)+2\hat{h}_n(Y_2)+\hat{f}_n(Y_2)\right),\\
2i\pi T^2&=&-\log{Y_2}+2\sum_{n=1}^\infty \frac{(2n-1)!}{(n!)^2}Y_2^{-n}-\frac{1}{\varpi_0}\sum_{n=1}^\infty\frac{(6n)! (-1)^n}{(n!)^3 (3n)!}Y_1^{-n}\hat{f}_n(Y_2)~,
\end{eqnarray}
where besides the fundamental period the following auxiliary functions were used:
\begin{eqnarray}
&& Y_1=\frac{(12\psi)^6}{2\phi}~,\\
\nonumber && Y_2= (2\phi)^2~,\\
\nonumber && B_n=6\Psi(6n+1)-3\Psi(3n+1)-3\Psi(n+1)~,\\
\nonumber &&u_n=(2\phi)^n \hat{u}_n~,\\
\nonumber &&h_n=2 (2\phi)^n \hat{h}_n~,\\
\nonumber &&f_n=-\frac{(2\phi)^n}{\sqrt{\phi^2-1}}\hat{f}_n~.
\end{eqnarray}

\section{Decay constants in the multi-moduli setup}
\label{app:decayconstants}
To understand the decay constants in the multi-moduli case, let us start with the one modulus case which can be described by the following Lagrangian
\begin{equation}
{\cal L}=f_a^2 \partial_\mu \phi \partial^\mu \bar{\phi}-\Lambda^4 (1-\cos{(2\pi \alpha\phi)})~.
\end{equation}
By going to canonical kinetic terms $(\hat{\phi}= f_a \phi)$ the Lagrangian becomes
\begin{equation}
{\cal L}= \partial_\mu \hat{\phi} \partial^\mu \bar{\hat{\phi}}-\Lambda^4 \left[1-\cos{\left(\frac{2\pi\alpha\hat{\phi}}{f_a}\right)}\right].
\end{equation}
In the main text, we have not specified the potential but restricted ourselves to the situation where $\alpha=1$ as the potential then obeys the shift symmetry $\phi\to\phi+1.$ The physical fields then have a shift symmetry $\hat{\phi}\to \hat{\phi}+f_a.$ The quadratic term of this field expanded around the origin is
\begin{equation}
4\pi^2\Lambda^4\frac{\hat{\phi}^2 }{f_a^2}~.
\label{eq:onemodulusmass}
\end{equation}

\begin{figure}
\includegraphics[width=0.28\textwidth]{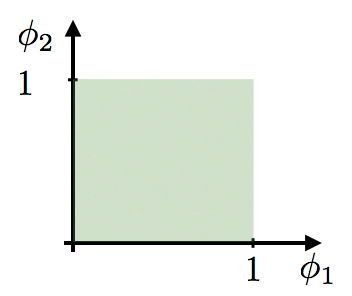}\includegraphics[width=0.3\textwidth]{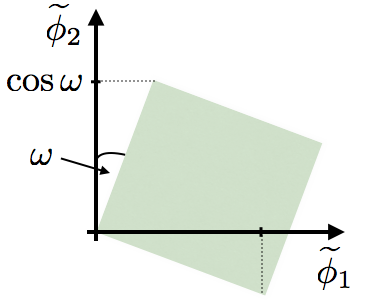}\includegraphics[width=0.35\textwidth]{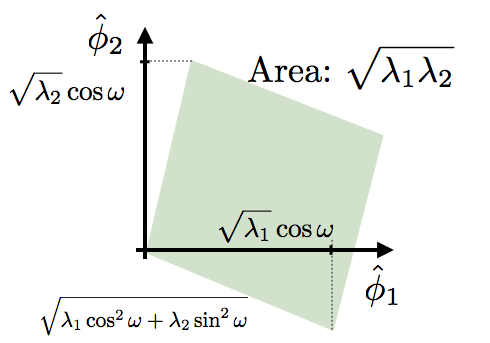}
\caption{Left: Fundamental domain in the original coordinates. We have a shift symmetry $\phi\to\phi+1.$ Middle: After diagonalising the kinetic terms, the fundamental domain is rotated. We denote the rotation angle by $\omega.$ Right: Going to canonical kinetic terms, the physical area of the fundamental domain is given by $\sqrt{\lambda_1\lambda_2}.$ The fundamental domain becomes a parallelogram.}
\label{fig:firstrotation}
\end{figure}

In the two (multi) moduli case, we proceed similarly, by demanding a shift symmetry of the potential $\phi_i\to\phi_i+1.$ A canonical example would be
\begin{equation}
{\cal L}=K_{i\bar{j}}\partial_\mu \phi\partial^\mu \bar{\phi}-\Lambda^4\sum_i (1-\cos{\left(2\pi\phi_i\right)})~.
\end{equation}
What is the shift symmetry for the physical fields? We first rotate to a basis where the kinetic terms are diagonal
\begin{equation}
\tilde{\phi}_i=R_{ij}\phi_j~,
\end{equation}
where $R_{ij}\in SO(2)$ and is given by the normalised eigenvectors of the K\"ahler metric.
\begin{equation}
{\cal L}=\lambda_i |\partial\tilde{\phi}_i|^2-\Lambda^4\sum_i \left[1-\cos{\left(2\pi R_{ij}^T\tilde{\phi}_j\right)}\right],
\end{equation}
where $\lambda_i$ are the eigenvalues of the K\"ahler metric.
The shift symmetry is illustrated in Figure~\ref{fig:firstrotation}. In the second step we re-scale the kinetic terms such that they become canonical $\hat{\phi}_i=\sqrt{\lambda_i}\tilde{\phi_i}.$ This the analogous step to the transformation in the one-modulus case. The Lagrangian becomes
\begin{equation}
{\cal L}=|\partial\hat{\phi}_i|^2-\Lambda^4\sum_i \left[1-\cos{\left(2\pi R_{ij}^T\frac{\hat{\phi}_j}{\sqrt{\lambda_j}}\right)}\right].
\end{equation}
The mass term is diagonal and is given by
\begin{equation}
4\pi^2\Lambda^4\left(\frac{\hat{\phi}_1^2}{\lambda_1}+\frac{\hat{\phi}_2^2}{\lambda_2}\right).
\end{equation}
This term is in complete analogy to the mass term in the one modulus case~\eqref{eq:onemodulusmass}.
This is the situation of two axions with decay constants $\sqrt{\lambda_{1}}$ and $\sqrt{\lambda_{2}}$ respectively.

\bibliography{axionbib2}{}

\providecommand{\href}[2]{#2}\begingroup\raggedright\begin{thebibliography}{10}

\bibitem{Freese:1990rb}
K.~Freese, J.~A. Frieman, and A.~V. Olinto, {\it {Natural inflation with pseudo
  - Nambu-Goldstone bosons}},  {\em Phys. Rev. Lett.} {\bf 65} (1990)
  3233--3236.

\bibitem{Kim:2004rp}
J.~E. Kim, H.~P. Nilles, and M.~Peloso, {\it {Completing natural inflation}},
  {\em JCAP} {\bf 0501} (2005) 005,
  [\href{http://xxx.lanl.gov/abs/hep-ph/0409138}{{\tt hep-ph/0409138}}].

\bibitem{Dimopoulos:2005ac}
S.~Dimopoulos, S.~Kachru, J.~McGreevy, and J.~G. Wacker, {\it {N-flation}},
  {\em JCAP} {\bf 0808} (2008) 003,
  [\href{http://xxx.lanl.gov/abs/hep-th/0507205}{{\tt hep-th/0507205}}].

\bibitem{Silverstein:2008sg}
E.~Silverstein and A.~Westphal, {\it {Monodromy in the CMB: Gravity Waves and
  String Inflation}},  {\em Phys. Rev.} {\bf D78} (2008) 106003,
  [\href{http://xxx.lanl.gov/abs/0803.3085}{{\tt 0803.3085}}].

\bibitem{McAllister:2008hb}
L.~McAllister, E.~Silverstein, and A.~Westphal, {\it {Gravity Waves and Linear
  Inflation from Axion Monodromy}},  {\em Phys. Rev.} {\bf D82} (2010) 046003,
  [\href{http://xxx.lanl.gov/abs/0808.0706}{{\tt 0808.0706}}].

\bibitem{Cicoli:2008gp}
M.~Cicoli, C.~P. Burgess, and F.~Quevedo, {\it {Fibre Inflation: Observable
  Gravity Waves from IIB String Compactifications}},  {\em JCAP} {\bf 0903}
  (2009) 013, [\href{http://xxx.lanl.gov/abs/0808.0691}{{\tt 0808.0691}}].

\bibitem{Kaloper:2008fb}
N.~Kaloper and L.~Sorbo, {\it {A Natural Framework for Chaotic Inflation}},
  {\em Phys. Rev. Lett.} {\bf 102} (2009) 121301,
  [\href{http://xxx.lanl.gov/abs/0811.1989}{{\tt 0811.1989}}].

\bibitem{Berg:2009tg}
M.~Berg, E.~Pajer, and S.~Sjors, {\it {Dante's Inferno}},  {\em Phys. Rev.}
  {\bf D81} (2010) 103535, [\href{http://xxx.lanl.gov/abs/0912.1341}{{\tt
  0912.1341}}].

\bibitem{Palti:2014kza}
E.~Palti and T.~Weigand, {\it {Towards large r from [p, q]-inflation}},  {\em
  JHEP} {\bf 04} (2014) 155, [\href{http://xxx.lanl.gov/abs/1403.7507}{{\tt
  1403.7507}}].

\bibitem{Marchesano:2014mla}
F.~Marchesano, G.~Shiu, and A.~M. Uranga, {\it {F-term Axion Monodromy
  Inflation}},  {\em JHEP} {\bf 09} (2014) 184,
  [\href{http://xxx.lanl.gov/abs/1404.3040}{{\tt 1404.3040}}].

\bibitem{Blumenhagen:2014gta}
R.~Blumenhagen and E.~Plauschinn, {\it {Towards Universal Axion Inflation and
  Reheating in String Theory}},  {\em Phys. Lett.} {\bf B736} (2014) 482--487,
  [\href{http://xxx.lanl.gov/abs/1404.3542}{{\tt 1404.3542}}].

\bibitem{Hebecker:2014eua}
A.~Hebecker, S.~C. Kraus, and L.~T. Witkowski, {\it {D7-Brane Chaotic
  Inflation}},  {\em Phys. Lett.} {\bf B737} (2014) 16--22,
  [\href{http://xxx.lanl.gov/abs/1404.3711}{{\tt 1404.3711}}].

\bibitem{Ibanez:2014kia}
L.~E. Ib\'a\~nez and I.~Valenzuela, {\it {The inflaton as an MSSM Higgs and
  open string modulus monodromy inflation}},  {\em Phys. Lett.} {\bf B736}
  (2014) 226--230, [\href{http://xxx.lanl.gov/abs/1404.5235}{{\tt 1404.5235}}].

\bibitem{Ben-Dayan:2014lca}
I.~Ben-Dayan, F.~G. Pedro, and A.~Westphal, {\it {Towards Natural Inflation in
  String Theory}},  {\em Phys. Rev.} {\bf D92} (2015), no.~2 023515,
  [\href{http://xxx.lanl.gov/abs/1407.2562}{{\tt 1407.2562}}].

\bibitem{Baumann:2014nda}
D.~Baumann and L.~McAllister, {\em {Inflation and String Theory}}.
\newblock Cambridge University Press, 2015.

\bibitem{Westphal:2015eva}
A.~Westphal, {\it {String Cosmology -- Large-Field Inflation in String
  Theory}},  {\em Adv. Ser. Direct. High Energy Phys.} {\bf 22} (2015)
  351--384.

\bibitem{Banks:2003sx}
T.~Banks, M.~Dine, P.~J. Fox, and E.~Gorbatov, {\it {On the possibility of
  large axion decay constants}},  {\em JCAP} {\bf 0306} (2003) 001,
  [\href{http://xxx.lanl.gov/abs/hep-th/0303252}{{\tt hep-th/0303252}}].

\bibitem{Baumann:2006cd}
D.~Baumann and L.~McAllister, {\it {A Microscopic Limit on Gravitational Waves
  from D-brane Inflation}},  {\em Phys. Rev.} {\bf D75} (2007) 123508,
  [\href{http://xxx.lanl.gov/abs/hep-th/0610285}{{\tt hep-th/0610285}}].

\bibitem{Conlon:2012tz}
J.~P. Conlon, {\it {Quantum Gravity Constraints on Inflation}},  {\em JCAP}
  {\bf 1209} (2012) 019, [\href{http://xxx.lanl.gov/abs/1203.5476}{{\tt
  1203.5476}}].

\bibitem{Kaloper:2015jcz}
N.~Kaloper, M.~Kleban, A.~Lawrence, and M.~S. Sloth, {\it {Large Field
  Inflation and Gravitational Entropy}},
  \href{http://xxx.lanl.gov/abs/1511.05119}{{\tt 1511.05119}}.

\bibitem{Conlon:2011qp}
J.~P. Conlon, {\it {Brane-Antibrane Backreaction in Axion Monodromy
  Inflation}},  {\em JCAP} {\bf 1201} (2012) 033,
  [\href{http://xxx.lanl.gov/abs/1110.6454}{{\tt 1110.6454}}].

\bibitem{Palti:2015xra}
E.~Palti, {\it {On Natural Inflation and Moduli Stabilisation in String
  Theory}},  {\em JHEP} {\bf 10} (2015) 188,
  [\href{http://xxx.lanl.gov/abs/1508.00009}{{\tt 1508.00009}}].

\bibitem{ArkaniHamed:2006dz}
N.~Arkani-Hamed, L.~Motl, A.~Nicolis, and C.~Vafa, {\it {The String landscape,
  black holes and gravity as the weakest force}},  {\em JHEP} {\bf 06} (2007)
  060, [\href{http://xxx.lanl.gov/abs/hep-th/0601001}{{\tt hep-th/0601001}}].

\bibitem{delaFuente:2014aca}
A.~de~la Fuente, P.~Saraswat, and R.~Sundrum, {\it {Natural Inflation and
  Quantum Gravity}},  {\em Phys. Rev. Lett.} {\bf 114} (2015), no.~15 151303,
  [\href{http://xxx.lanl.gov/abs/1412.3457}{{\tt 1412.3457}}].

\bibitem{Brown:2015iha}
J.~Brown, W.~Cottrell, G.~Shiu, and P.~Soler, {\it {Fencing in the Swampland:
  Quantum Gravity Constraints on Large Field Inflation}},  {\em JHEP} {\bf 10}
  (2015) 023, [\href{http://xxx.lanl.gov/abs/1503.04783}{{\tt 1503.04783}}].

\bibitem{Montero:2015ofa}
M.~Montero, A.~M. Uranga, and I.~Valenzuela, {\it {Transplanckian axions!?}},
  {\em JHEP} {\bf 08} (2015) 032,
  [\href{http://xxx.lanl.gov/abs/1503.03886}{{\tt 1503.03886}}].

\bibitem{Bachlechner:2015qja}
T.~C. Bachlechner, C.~Long, and L.~McAllister, {\it {Planckian Axions and the
  Weak Gravity Conjecture}},  \href{http://xxx.lanl.gov/abs/1503.07853}{{\tt
  1503.07853}}.

\bibitem{Hebecker:2015rya}
A.~Hebecker, P.~Mangat, F.~Rompineve, and L.~T. Witkowski, {\it {Winding out of
  the Swamp: Evading the Weak Gravity Conjecture with F-term Winding
  Inflation?}},  {\em Phys. Lett.} {\bf B748} (2015) 455--462,
  [\href{http://xxx.lanl.gov/abs/1503.07912}{{\tt 1503.07912}}].

\bibitem{Brown:2015lia}
J.~Brown, W.~Cottrell, G.~Shiu, and P.~Soler, {\it {On Axionic Field Ranges,
  Loopholes and the Weak Gravity Conjecture}},
  \href{http://xxx.lanl.gov/abs/1504.00659}{{\tt 1504.00659}}.

\bibitem{Rudelius:2015xta}
T.~Rudelius, {\it {Constraints on Axion Inflation from the Weak Gravity
  Conjecture}},  {\em JCAP} {\bf 1509} (2015), no.~09 020,
  [\href{http://xxx.lanl.gov/abs/1503.00795}{{\tt 1503.00795}}].

\bibitem{Junghans:2015hba}
D.~Junghans, {\it {Large-Field Inflation with Multiple Axions and the Weak
  Gravity Conjecture}},  \href{http://xxx.lanl.gov/abs/1504.03566}{{\tt
  1504.03566}}.

\bibitem{Heidenreich:2015wga}
B.~Heidenreich, M.~Reece, and T.~Rudelius, {\it {Weak Gravity Strongly
  Constrains Large-Field Axion Inflation}},
  \href{http://xxx.lanl.gov/abs/1506.03447}{{\tt 1506.03447}}.

\bibitem{Heidenreich:2015nta}
B.~Heidenreich, M.~Reece, and T.~Rudelius, {\it {Sharpening the Weak Gravity
  Conjecture with Dimensional Reduction}},
  \href{http://xxx.lanl.gov/abs/1509.06374}{{\tt 1509.06374}}.

\bibitem{Kooner:2015rza}
K.~Kooner, S.~Parameswaran, and I.~Zavala, {\it {Warping the Weak Gravity
  Conjecture}},  \href{http://xxx.lanl.gov/abs/1509.07049}{{\tt 1509.07049}}.

\bibitem{Hebecker:2015zss}
A.~Hebecker, F.~Rompineve, and A.~Westphal, {\it {Axion Monodromy and the Weak
  Gravity Conjecture}},  \href{http://xxx.lanl.gov/abs/1512.03768}{{\tt
  1512.03768}}.

\bibitem{1401.2579}
M.~Cicoli, K.~Dutta, and A.~Maharana, {\it {N-flation with Hierarchically Light
  Axions in String Compactifications}},  {\em JCAP} {\bf 1408} (2014) 012,
  [\href{http://xxx.lanl.gov/abs/1401.2579}{{\tt 1401.2579}}].

\bibitem{Witten:1981gj}
E.~Witten, {\it {Instability of the Kaluza-Klein Vacuum}},  {\em Nucl. Phys.}
  {\bf B195} (1982) 481.

\bibitem{RoblesLlana:2006ez}
D.~Robles-Llana, F.~Saueressig, and S.~Vandoren, {\it {String loop corrected
  hypermultiplet moduli spaces}},  {\em JHEP} {\bf 03} (2006) 081,
  [\href{http://xxx.lanl.gov/abs/hep-th/0602164}{{\tt hep-th/0602164}}].

\bibitem{Candelas:1990rm}
P.~Candelas, X.~C. De~La~Ossa, P.~S. Green, and L.~Parkes, {\it {A Pair of
  Calabi-Yau manifolds as an exactly soluble superconformal theory}},  {\em
  Nucl. Phys.} {\bf B359} (1991) 21--74.

\bibitem{Klemm:1992tx}
A.~Klemm and S.~Theisen, {\it {Considerations of one modulus Calabi-Yau
  compactifications: Picard-Fuchs equations, Kahler potentials and mirror
  maps}},  {\em Nucl. Phys.} {\bf B389} (1993) 153--180,
  [\href{http://xxx.lanl.gov/abs/hep-th/9205041}{{\tt hep-th/9205041}}].

\bibitem{Klemm:1993jj}
A.~Klemm and S.~Theisen, {\it {Mirror maps and instanton sums for complete
  intersections in weighted projective space}},  {\em Mod. Phys. Lett.} {\bf
  A9} (1994) 1807--1818, [\href{http://xxx.lanl.gov/abs/hep-th/9304034}{{\tt
  hep-th/9304034}}].

\bibitem{Kawada:2015vwe}
H.~Kawada, T.~Masuda, and H.~Suzuki, {\it {Mirror Symmetry of Minimal
  Calabi-Yau Manifolds}},  \href{http://xxx.lanl.gov/abs/1512.07737}{{\tt
  1512.07737}}.

\bibitem{Braun:2015jdy}
V.~Braun, P.~Candelas, and X.~de~la Ossa, {\it {Two One-Parameter Special
  Geometries}},  \href{http://xxx.lanl.gov/abs/1512.08367}{{\tt 1512.08367}}.

\bibitem{Braun:2009qy}
V.~Braun, P.~Candelas, and R.~Davies, {\it {A Three-Generation Calabi-Yau
  Manifold with Small Hodge Numbers}},  {\em Fortsch. Phys.} {\bf 58} (2010)
  467--502, [\href{http://xxx.lanl.gov/abs/0910.5464}{{\tt 0910.5464}}].

\bibitem{Candelas:1993dm}
P.~Candelas, X.~De~La~Ossa, A.~Font, S.~H. Katz, and D.~R. Morrison, {\it
  {Mirror symmetry for two parameter models. 1.}},  {\em Nucl. Phys.} {\bf
  B416} (1994) 481--538, [\href{http://xxx.lanl.gov/abs/hep-th/9308083}{{\tt
  hep-th/9308083}}].

\bibitem{Giryavets:2003vd}
A.~Giryavets, S.~Kachru, P.~K. Tripathy, and S.~P. Trivedi, {\it {Flux
  compactifications on Calabi-Yau threefolds}},  {\em JHEP} {\bf 04} (2004)
  003, [\href{http://xxx.lanl.gov/abs/hep-th/0312104}{{\tt hep-th/0312104}}].

\bibitem{Kappl:2014lra}
R.~Kappl, S.~Krippendorf, and H.~P. Nilles, {\it {Aligned Natural Inflation:
  Monodromies of two Axions}},  {\em Phys. Lett.} {\bf B737} (2014) 124--128,
  [\href{http://xxx.lanl.gov/abs/1404.7127}{{\tt 1404.7127}}].

\bibitem{Burgess:2014oma}
C.~Burgess and D.~Roest, {\it {Inflation by Alignment}},  {\em JCAP} {\bf 1506}
  (2015), no.~06 012, [\href{http://xxx.lanl.gov/abs/1412.1614}{{\tt
  1412.1614}}].

\bibitem{Long:2014dta}
C.~Long, L.~McAllister, and P.~McGuirk, {\it {Aligned Natural Inflation in
  String Theory}},  {\em Phys. Rev.} {\bf D90} (2014) 023501,
  [\href{http://xxx.lanl.gov/abs/1404.7852}{{\tt 1404.7852}}].

\bibitem{Choi:2014rja}
K.~Choi, H.~Kim, and S.~Yun, {\it {Natural inflation with multiple
  sub-Planckian axions}},  {\em Phys. Rev.} {\bf D90} (2014) 023545,
  [\href{http://xxx.lanl.gov/abs/1404.6209}{{\tt 1404.6209}}].

\bibitem{Long:2014fba}
C.~Long, L.~McAllister, and P.~McGuirk, {\it {Heavy Tails in Calabi-Yau Moduli
  Spaces}},  {\em JHEP} {\bf 10} (2014) 187,
  [\href{http://xxx.lanl.gov/abs/1407.0709}{{\tt 1407.0709}}].

\bibitem{Kappl:2015esy}
R.~Kappl, H.~P. Nilles, and M.~W. Winkler, {\it {Modulated Natural Inflation}},
   \href{http://xxx.lanl.gov/abs/1511.05560}{{\tt 1511.05560}}.

\bibitem{Choi:2015aem}
K.~Choi and H.~Kim, {\it {Aligned Natural Inflation with Modulations}},
  \href{http://xxx.lanl.gov/abs/1511.07201}{{\tt 1511.07201}}.

\end{thebibliography}\endgroup
\bibliographystyle{JHEP}
\end{document}